# Exploring quantum interference in heteroatom-substituted graphene-like molecules

Sara Sangtarash[*], Hatef Sadeghi, and Colin J. Lambert[*]

Quantum Technology Centre, Physics Department, Lancaster University, Lancaster LA1 4YB, UK

[*]s.sangtarash@lancaster.ac.uk; [*]c.lambert@lancaster.ac.uk

If design principles for controlling quantum interference in single molecules could be elucidated and verified, then this will lay the foundations for exploiting such effects in nanoscale devices and thin-film materials. When the core of a graphene-like polyaromatic hydrocarbon (PAH) is weakly coupled to external electrodes by atoms $i$ and $j$, the single-molecule electrical conductance $\sigma_{ij}$ depends on the choice of connecting atoms $i, j$. Furthermore, conductance ratios $\sigma_{ij}/\sigma_{lm}$ corresponding to different connectivities $i, j$ and $l,m$ are determined by quantum interference within the PAH core. In this paper, we examine how such conductance ratios change when one of the carbon atoms within the 'parent' PAH core is replaced by a heteroatom to yield a 'daughter' molecule. For bipartite parental cores, in which odd-numbered sites are connected to even-numbered sites only, the effect of heteroatom substitution onto an odd-numbered site is summarized by the following qualitative rules:

(a) When $i$ and $j$ are odd, both parent and daughter have low conductances

(b) When $i$ is odd and $j$ is even, or vice versa both parent and daughter have high conductances

(c) When $i,j$ are both even, the parent has a low conductance and the daughter a high conductance.

These rules are verified by comparison with density-functional calculations on naphthalene, anthracene, pyrene and anthanthrene cores connected via two different anchor groups to gold electrodes.

Single-molecule electronics is a branch of fundamental science, which aims to probe the transport of electrons through molecular junctions formed from single molecules connected to source and drain electrodes[1]. Ultimately, such studies will inform the discovery of new materials and devices with unprecedented properties and performance, but first their design rules need to be clarified[2,3]. In practice this challenge has to be met within the context of measurement variability and limitations in current theories, because all techniques for measuring the electrical conductance $G$ of single molecules[4-20] yield only statistical distributions of $G$, while material-specific simulation tools such as density functional theory (DFT), GW many-body theory and non-equilibrium Green's functions (NEGFs) require free parameters to adjust energy gaps[21] and energies of frontier orbitals relative to the Fermi energy $E_F$ of the electrodes[2,22]. This means that such simulation tools can be used to post-rationalise experiments and predict trends, but they cannot be relied upon to predict quantitatively the properties of any particular junction.

Within this landscape of uncertainties, analytic theories play an important role, because they identify generic behaviours, which are potentially common to all molecules. One example is on-resonance constructive interference captured by the Breit-Wigner formula[2], which applies when the Fermi energy is close to either the HOMO or LUMO levels and states that the transmission coefficient $T(E)$ describing electrons of energy $E$ passing through the molecule from one electrode to the other should have a Lorentzian dependence on $E$. Similarly generic features of destructive interference are described by transmission functions with Fano lineshapes[23]. On the other hand, unless a molecule is externally gated, the Fermi energy is usually located within the HOMO-LUMO gap and transport takes place via off-resonance, phase-coherent co-tunnelling[2-4,20]. In this case, for polyaromatic hydrocarbons (PAHs) in the co-tunnelling regime, counting rules[24-28] can be used to identify conditions for the occurrence of *destructive* quantum interference, while our recently-developed mid-gap theory and magic numbers[3,4] can be used to account for *constructive* interference. Furthermore for molecules whose central core is weakly coupled to external electrodes, quantum circuit rules[29,30] allow measurements of molecules of the form X-C-X and Y-C-Y to predict properties of molecules of the form X-C-Y.

Beyond merely describing QI in single molecules, it is desirable to identify a conceptual framework for controlling QI, particularly through chemical modification. For example for the purpose of increasing the electrical conductance of single molecules and molecular films, it is desirable to identify strategies for removing destructive QI. In this paper we report a new conceptual advance in the scientific understanding and technological know-how necessary to control quantum effects in single molecules, by formulating three rules which answer the question: How does heteroatom substitution affect single-molecule electrical conductance? Our aim is to examine the effect of heteroatom substitution on electron transport through PAHs. We find that the conductance of PAHs exhibiting destructive interference can be significantly increased by the introduction of a heteroatom, whereas the conductance PAHs exhibiting constructive interference are barely affected by heteroatom substitution.

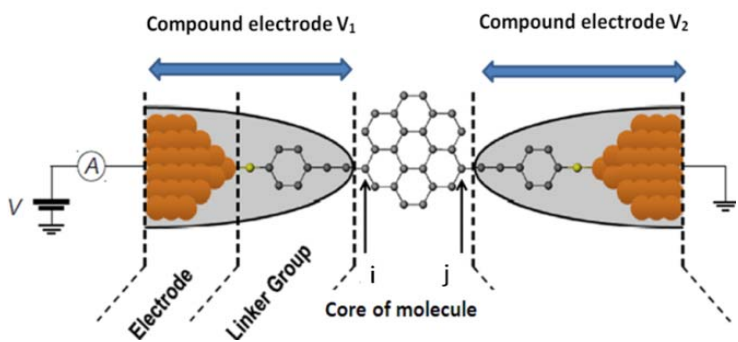

Figure 1. A PAH core, weakly coupled to 'compound electrodes'.

Figure 1 shows an example of a PAH, whose central aromatic core can be regarded as weakly connected to 'compound electrodes' via sites $i$ and $j$. In this example, each compound electrode is composed of a linker group comprising a triple bond connected to a phenyl ring, which in turn is connected to a gold electrode via a sulfur anchor atom. In refs[2-4], we noted that provided the compound electrodes are weakly coupled to the central core, the electrical conductance $\boldsymbol{\sigma_{ij}}$ of such a junction can be written as a product of the form

$$\sigma_{ij} = V_1 \tau_{ij} V_2, \qquad (1)$$

where the connectivity-independent terms $V_1$ and $V_2$ are associated with the compound electrodes, while $\tau_{ij}$ is a contribution from the central core, which depends on the connectivity $i,j$. The validity of equation (1) requires that the Fermi energy of the gold should lie within the HOMO-LUMO gap of the central core, so that transport takes place via phase-coherent co-tunneling and multiple scattering effects between the core and virtual electrodes are suppressed[3]. Equation (1) is significant, because it implies (see mathematical methods) that ratios of experimentally-reported "statistically-most-probable conductances" corresponding to different connectivities $i,j$ and $l,m$ are properties of the core of the molecule and satisfy

$$\frac{\sigma_{ij}}{\sigma_{lm}} = \frac{\tau_{ij}}{\tau_{lm}} \qquad (2)$$

which allows us to predict conductance ratios from a knowledge of the core alone[3]. For PAH cores[3,4], we also demonstrated that the effect of connectivity on the transmissions $\tau_{ij}$ could be calculated by introducing tables of '*magic numbers*' $M_{ij}$, from which one obtains $\tau_{ij} = (M_{ij})^2$. Furthermore we showed that the conductance ratios predicted by this simple 'analytic M-theory' are in close agreement with experiment[3].

**The role of heteroatoms.** The aim of the present paper is to examine how connectivity-dependent conductance ratios change in the presence of heteroatoms and to elucidate simple rules which allow us to anticipate the effect of heteroatom substitution. Examples of lattices describing such molecules are shown in figure 2 below. In the absence of heteroatoms, the 'parent' PAH Hamiltonian $H$ of each of lattice is modelled by a connectivity matrix, with entries -1 connecting neighbouring sites and all other elements set to zero. Such PAHs are bipartite lattices with a filled HOMO and empty LUMO, in which odd-numbered sites are connected to even-numbered sites only and vice versa. Consequently the magic numbers $M_{ij}^p$ of such parent lattices are simply integers. When $i$ is even and $j$ is odd (or vice versa), $M_{ij}^p$ is non-zero, corresponding to connectivities exhibiting constructive interference within the middle of the HOMO-LUMO gap of the core of the molecule. In contrast $M_{ij}^p$ vanishes when $i$ and $j$ are both even or both odd and odd, reflecting the fact that these connectivities correspond to destructive interference at the gap centre.

To each of the above parent lattices, we now examine the properties of the corresponding 'daughter,' obtained by substituting a single heteroatom on a site $l$, whose effect is modelled by adding a single non-zero diagonal element $H_{ll}$ to the Hamiltonian. Without loss of generality, we choose $l$ to be an odd-numbered site $l=1$.

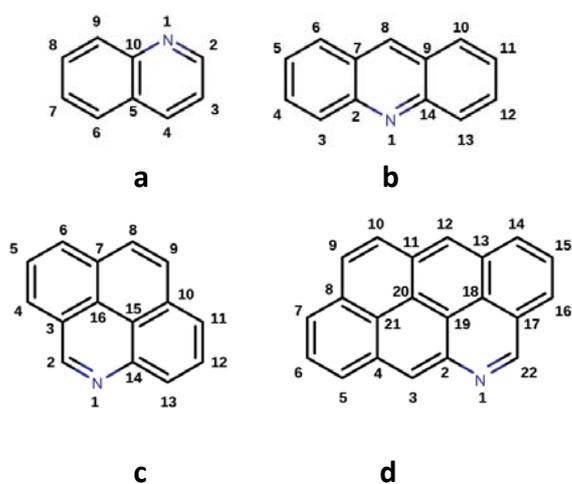

**Figure 2.** Molecular structures of substituted (a) naphthalene (b) anthracene (c) pyrene (d) anthanthrene, with a hetroatom in site 1.

Our main finding is that provided the daughter and parent share the same Fermi energy (corresponding to the mid-gap of the parent), the magic numbers $M_{ij}^d$ of such daughter lattices satisfy three rules (for detailed description see mathematical methods):

(1) When both of $i,j$ are odd, $M_{ij}^d = M_{ij}^p = 0$ and therefore for odd-to-odd connectivities, quantum interference remains destructive for both the parent and daughter, and is unaffected by the presence of the heteroatom. Qualitatively in a real molecule, this means that when $i$ and $j$ are odd, both parent and daughter have low conductances.

(2) When $i$ is odd and $j$ is even, or vice versa, $M_{ij}^d = M_{ij}^p \neq 0$ and therefore interference is constructive for both the parent and daughter, and is unaffected by the heteroatom. Qualitatively, this means that when $i$ is odd and $j$ is even, or vice versa both parent and daughter have high conductances. This resilience of constructive interference to the presence of a heteroatom occurs for example in a benzene ring, when $i$ and $j$ correspond to para connectivities of the parent and was observed recently in ref [31] for a benzene parent and pyridine daughter, where the presence of a heteroatom was found to have little effect.

(3) When $i,j$ are both even $M_{ij}^d = cM_{il}^p M_{lj}^p$, where c is a connectivity-independent constant. Therefore the parent suffers destructive interference, whereas for the daughter, interference becomes constructive. From the viewpoint of conductances, this implies that when $i,j$ are both even, the parent has a low conductance and the daughter a high conductance.

In summary, at the mid-gap of the parent, the addition of a heteroatom to the odd sub-lattice removes the destructive interference associated with even-to-even connectivities of the parent, but has a much smaller effect on other connectivitites. This holds not only at the mid-gap of the parent, but also in the vicinity of the middle of the HOMO-LUMO gap as shown below. Interestingly, from equation (2) and rule 3, provided the daughter and parent share the same Fermi energy, the ratio of two of the resulting daughter conductances $\sigma_{ij}^d$, $\sigma_{kq}^d$ associated with two different even-to-even connectivities $i,j$ and $k,q$ is independent of the hetero-atom parameter $c$ and takes the form:

$$\frac{\sigma_{ij}^d}{\sigma_{kq}^d} = \frac{\tau_{ij}^d}{\tau_{kq}^d} = \left(\frac{M_{il}^p M_{lj}^p}{M_{kl}^p M_{lq}^p}\right)^2 = \left(\frac{\tau_{il}^p \tau_{lj}^p}{\tau_{kl}^p \tau_{lq}^p}\right) = \left(\frac{\sigma_{il}^p \sigma_{lj}^p}{\sigma_{kl}^p \sigma_{lq}^p}\right) \qquad (3)$$

Hence, measurement of the four parental conductances on the right hand side of equation (3) allows us to predict the ratio of the two daughter conductances on the left.

**Comparison between analytic M-theory and density functional theory (DFT) for parent and daughter conductance ratios.** To demonstrate that the above M-theory rules are a useful guideline for anticipating the effect of heteroatom substitution on the electrical conductance of the above PAHs, we now make a comparison with DFT calculations of the transmission coefficients $T_{ij}(E)$ describing electrons passing through the core of a molecule from one compound electrode to the other. Figure 3 shows examples of PAHs attached to "compound electrodes" comprising acetylene linkers attached via phenyl rings and a thiol anchor to gold electrodes, for which we use the Gollum transport code[32] to compute $T_{ij}(E)$ for various connectivities $i,j$.

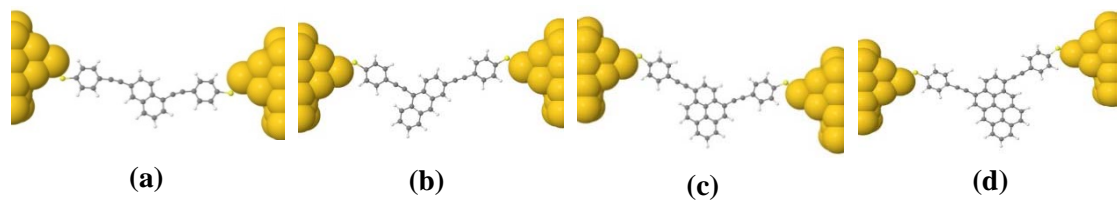

(a)        (b)        (c)        (d)

**Figure 3**. Examples of molecules placed between gold electrodes (a) naphthalene, (b) anthracene, (c) pyrene and (d) anthanthrene

As discussed in the mathematical methods and ref[3], to each magic number $M_{ij}$ there is an energy-dependent M-function $M_{ij}(E)$ (see SI), and a corresponding energy-dependent core transmission function $\tau_{ij}(E) = M_{ij}(E)^2$. The quantities $\tau_{ij}^d$ and $\tau_{ij}^p$ appearing in equation (3) are simply the values of their daughter and parent transmission functions evaluated at $E=0$. If either $i$ or $j$ are odd, then rule 2 yields $\tau_{ij}^d(0) = \tau_{ij}^p(0)$.

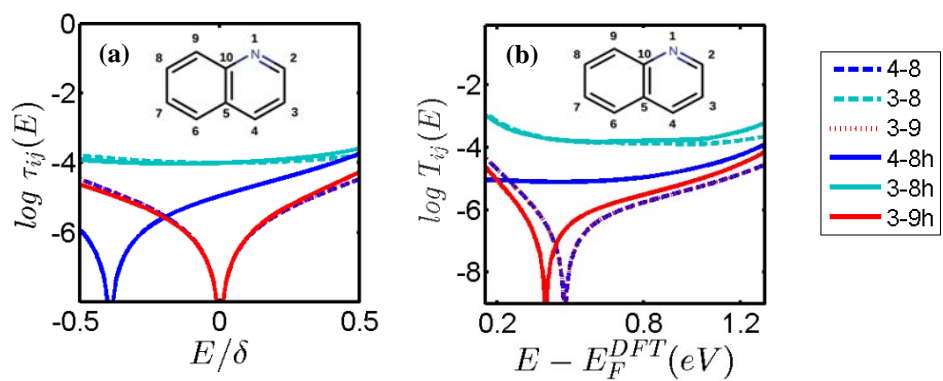

**Figure 4** (a) Core transmission coefficients $\tau_{i,j}(E)$ of naphthalene parents (dashed lines) and quinolone daughters (solid lines) plotted against $E/\delta$, where $\delta$ is half of the HOMO-LUMO gap of the parental core; ie $\delta=0.62$ for naphthalene. (b) *DFT* results for the corresponding the transmission coefficients $T_{ij}(E)$ of the molecules connected to gold electrodes.

Figure 4a shows the core transmission functions $\tau_{ij}^d(E)$ (solid lines) and $\tau_{ij}^p(E)$ (dashed lines) for naphthalene and quinoline, respectively. For each quinoline daughter, the heteroatom is placed on site *l=1*. The 3-9h red solid curve of figures 4a and the almost-identical 3-9 red dotted parental curve, demonstrate that for odd-odd connectivities, destructive interference is present for both the parent and daughter and therefore rule 1 is satisfied. To illustrate rule 2, the 3-8 and 3-8h magenta curves of figure 4a demonstrate that the M-function and core transmission of the 3-8 para-connected parent are non-zero at the parent mid-gap energy (*E=0*), corresponding to constructive interference, and are unaffected by heteroatom substitution. Finally rule 3 is illustrated by the blue-dashed curve, which demonstrates that the core transmission of the 4-8 connected parent exhibits destructive interference, while in the vicinity of the mid-gap, the 4-8h solid-blue curve of the daughter becomes non-zero upon heteroatom substitution. To compare the above results with ab initio theory, figure 4b shows the transmission coefficients of these structures obtained from DFT (see Methods). The qualitative agreement between figures 4a and 4b demonstrates that connectivity plays a significant role in determining the outcome of the much more computationally-expensive DFT-based calculations. As shown in figures SI4 and SI5 of the SI, this conclusion holds also for the molecules with a different compound anchor formed from an acetylene linker attached to a gold electrode via a carbon-gold bond.

This qualitative agreement between M-theory and DFT is also evident in the results for the anthracene, pyrene and anthanthrene cores shown in figure 5, where figures 5(a-c) show the M-theory core transmissions for the parents and daughters, while figures 5(d-f) DFT results for the full transmission coefficients of the molecules attached to gold electrodes. M-tables for all of the above cores are presented in the SI, along with further comparisons between M-theory and DFT.

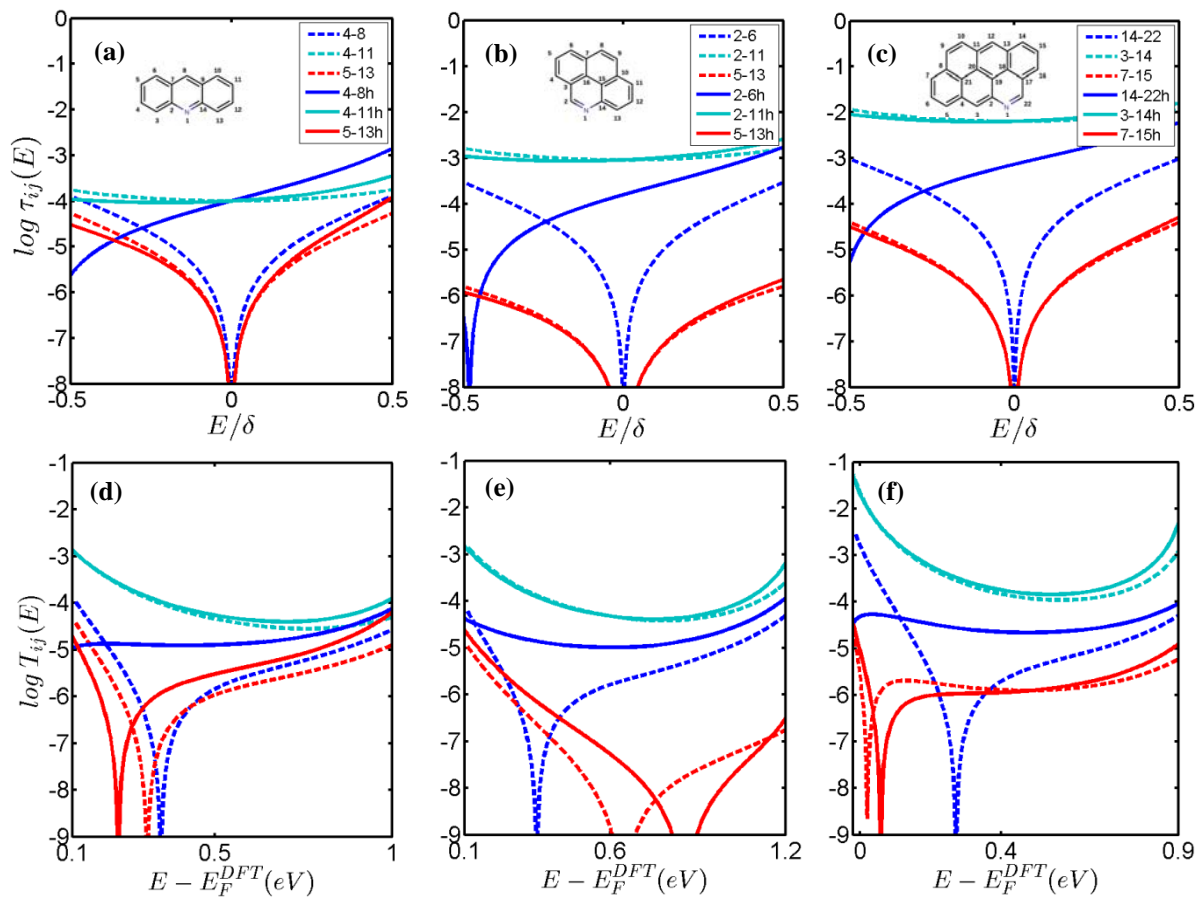

**Figure 5.** (a-c) Core transmissions of parents (dashed lines) and daughters (solid lines) ) plotted against E/δ, where δ is half of the HOMO-LUMO gap of the parental core. (d-f): DFT results for the transmission coefficients of daughters (solid lines) and parents (dashed lines) Parental cores are (a,d) Anthracene, (b,e) Pyrene and (c,f) Anthanthrene. For anthracene, pyrene and anthanthrene are δ=0.41, 0.45, 0.29 respectively.

A clear difference between M-theory and DFT is the location of the anti-resonances. M-theory places these at *E*=0, whereas DFT places anti-resonances at various energies relative to the DFT-predicted Fermi energy $E_F^{DFT}$. Nevertheless the similarities between the solid and dashed magenta lines shows that rule 2 is satisfied and rule 3 correctly predicts that heteroatom substitution creates a significant increase in conductance for the even-even connectivity. Finally the similarities between the solid and dashed red lines show that rule 1 correctly predicts low conductances for these connectivities, although in contrast with M-theory, the DFT-predicted energies for theparent and daughter anti-resonances are slightly different.

Since the DFT-predicted Fermi energy $E_F^{DFT}$ is not usually reliable, the correct value of Fermi energy $E_F$ relative to frontier orbitals is an unknown quantity and we consider it to be a free parameter. Equations (2) and (3) are valid when the Fermi energy $E_F^p$ of the parent and the Fermi energy $E_F^d$ of the

daughter are located at *E*=0 and are approximations otherwise. More generally, they should be replaced by

$$\frac{\sigma_{ij}^d}{\sigma_{kq}^d} = \frac{\tau_{ij}^d(E_F^d)}{\tau_{kq}^d(E_F^d)} \quad \text{and} \quad \frac{\sigma_{ij}^p}{\sigma_{kq}^p} = \frac{\tau_{ij}^p(E_F^p)}{\tau_{kq}^p(E_F^p)} \tag{4}$$

For the above PAHs, comparison with experiment suggest that $E_F^p = 0$, whereas currently there are no experiments which determine $E_F^d$. If the daughter Fermi energy differs from $E_F^d = 0$, then the above rules will be slightly modified and conductance ratios will differ from their $E_F^d = 0$ values. For example for the 2-8h and 2-6h connectivities of quinolene shown in figure SI3, the pink and green solid lines cross at *E*=0 and therefore if $E_F^d = 0$, we predict $\frac{\sigma_{28}^d}{\sigma_{26}^d} = \frac{\tau_{28}^d(0)}{\tau_{26}^d(0)} = 1$. On the other hand if $E_F^d$ takes a slightly positive value, we predict $\frac{\sigma_{28}^d}{\sigma_{26}^d} = \frac{\tau_{28}^d(E_F^d)}{\tau_{26}^d(E_F^d)} < 1$.

In summary, we have investigated the effect of heteroatom substitution on the electrical conductance of single-molecule junctions. For a bipartite PAH with a filled HOMO and empty LUMO, in which odd-numbered sites are connected only to even-numbered sites and vice versa, we found that provided the Fermi energy of the heteroatom-substituted daughter is close to the Fermi energy of the PAH parent, the addition of a heteroatom at site *l* only affects connectivities *i,j* if both of the parent connectivities *i,l* and *l,j* are constructive. This means that if *l* belongs to the odd-numbered sub-lattice, then the heteroatom alleviates the destructive interference associated with even-to-even connectivities of the parent and the change in conductance is described by qualitative three rules:

(a) When *i* and *j* are odd, both parent and daughter have low conductances

(b) When *i* is odd and *j* is even, or vice versa both parent and daughter have high conductances

(c) When *i,j* are both even, the parent has a low conductance and the daughter a high conductance.

The only available experiment investigating the effect of heteroatom substitution on the electrical conductance of PAHs[31] measured connectivities corresponding to constructive interference, and in agreement with rule b, found little effect. Consequently more experiments are needed to confirm the above predictions.

**Mathematical and Computational Methods**

**(a)    Ratios of statistically-most-probable conductances**

Equation (1) allows us to isolate the statistical fluctuations due to variability of the molecule-gold contacts from the properties of the core. For example, in experiments using mechanically-controlled break junctions, this variability is overcome by creating histograms of the logarithmic conductance $L_{ij}(V_1, V_2) = log_{10}\sigma_{ij}$ from thousands of conductance measurements and reporting the statistically-most probable value $\bar{L}_{ij}(V_1, V_2)$, or alternatively the conductance $\bar{\sigma}_{ij} = 10^{\bar{L}_{ij}(V_1,V_2)}$. Since this variability arises from fluctuations in $V_1$ and $V_2$, equation (1) yields

$$\bar{L}_{ij}(V_1, V_2) = log_{10}\bar{\sigma}_{ij} = log_{10}\bar{V}_1 + log_{10}\tau_{ij} + log_{10}\bar{V}_2 \qquad (5)$$

where $log_{10}\bar{V}_1$ is the statistically-most-probable value of $log_{10}V_1$ and similarly for $log_{10}\bar{V}_2$. As discussed in ref[30], equation (2) leads to 'quantum circuit rules' for electrical conductance and thermopower. Furthermore provided the statistics of the virtual electrodes is independent of connectivity, equation (2) yields

$$\frac{\bar{\sigma}_{ij}}{\bar{\sigma}_{lm}} = \frac{\tau_{ij}}{\tau_{lm}} \qquad (6)$$

This means that ratios of reported 'statistically-most-probable conductances' corresponding to different connectivities $i,j$ and $l,m$ are properties of the core of the molecule.

**(b)   M-theory for heteroatoms**

For simple PAH cores, to compute the quantities $\tau_{ij}$ a parameter-free theory of mid-gap electron transport (M-theory) was developed[3,4] in the weak coupling regime which describes the amplitude of the interference pattern on an arbitrary atomic orbital $i$ due to an electron of energy $E$ entering a core at orbital $j$. Each PAH is represented by a lattice of sites with nearest neighbour couplings and the Hamiltonian $H$ is equated to a simple connectivity table $C$, whose entries $C_{ij}$ were assigned a value -1 if sites $i$ and $j$ are nearest neighbours and a value of zero otherwise. The M-functions $M_{ij}(E)$ are then given by $i,j$ th elements of the matrix

$$M(E) = A(E)(E - C)^{-1} \qquad (7)$$

where $A(E)$ is a scalar function of $E$, chosen for convenience such that for parental PAH cores, $M_{ij}(0)$ is an integer. For a given core, $A(E)$ does not affect conductance ratios, because $\tau_{ij}(E) = (M_{ij}(E))^2$ and therefore $A(E)$ cancels in equation (2). When computing conductance ratios of parents and daughters, we choose the same $A(E)$ for both parent and daughter. Here, for simplicity, we choose $A(E)$ to be a constant, independent of energy (In what follows and in the SI, $A(E)$ is chosen to be 3, 4, 6 and 10 for naphthalene, anthracene, pyrene and anthanthrene, respectively). Rules 1-4 are a direct consequence of Dyson's equation, which for a heteroatom on site $l$ yields

$$M_{ij}^d(E) = M_{ij}^p(E) + M_{il}^p(E)cM_{lj}^p(E)/(1 - cM_{ll}^d(E)) \qquad (8)$$

For instance for the naphthalene core shown in figure 2b, the connectivity table $C^p$ of the parent molecule and the corresponding $M^p(0) = A(0)(-C^p)^{-1}$ table at the mid-gap of parent are constructed as:

| $C^p$ | 1 | 3 | 5 | 7 | 9 | 2 | 4 | 6 | 8 | 10 |
|---|---|---|---|---|---|---|---|---|---|---|
| 1 | 0 | 0 | 0 | 0 | 0 | -1 | 0 | 0 | 0 | -1 |
| 3 | 0 | 0 | 0 | 0 | 0 | -1 | -1 | 0 | 0 | 0 |
| 5 | 0 | 0 | 0 | 0 | 0 | 0 | -1 | -1 | 0 | -1 |
| 7 | 0 | 0 | 0 | 0 | 0 | 0 | 0 | -1 | -1 | 0 |
| 9 | 0 | 0 | 0 | 0 | 0 | 0 | 0 | 0 | -1 | -1 |
| 2 | -1 | -1 | 0 | 0 | 0 | 0 | 0 | 0 | 0 | 0 |
| 4 | 0 | -1 | -1 | 0 | 0 | 0 | 0 | 0 | 0 | 0 |
| 6 | 0 | 0 | -1 | -1 | 0 | 0 | 0 | 0 | 0 | 0 |
| 8 | 0 | 0 | 0 | -1 | -1 | 0 | 0 | 0 | 0 | 0 |
| 10 | -1 | 0 | -1 | 0 | -1 | 0 | 0 | 0 | 0 | 0 |

| $M^p$ | 1 | 3 | 5 | 7 | 9 | 2 | 4 | 6 | 8 | 10 |
|---|---|---|---|---|---|---|---|---|---|---|
| 1 | 0 | 0 | 0 | 0 | 0 | -2 | 2 | -1 | 1 | -1 |
| 3 | 0 | 0 | 0 | 0 | 0 | -1 | -2 | 1 | -1 | 1 |
| 5 | 0 | 0 | 0 | 0 | 0 | 1 | -1 | -1 | 1 | -1 |
| 7 | 0 | 0 | 0 | 0 | 0 | -1 | 1 | -2 | -1 | 1 |
| 9 | 0 | 0 | 0 | 0 | 0 | 1 | -1 | 2 | -2 | -1 |
| 2 | -2 | -1 | 1 | -1 | 1 | 0 | 0 | 0 | 0 | 0 |
| 4 | 2 | -2 | -1 | 1 | -1 | 0 | 0 | 0 | 0 | 0 |
| 6 | -1 | 1 | -1 | -2 | 2 | 0 | 0 | 0 | 0 | 0 |
| 8 | 1 | -1 | 1 | -1 | -2 | 0 | 0 | 0 | 0 | 0 |
| 10 | -1 | 1 | -1 | 1 | -1 | 0 | 0 | 0 | 0 | 0 |

By substituting a heteroatom with on-site energy of -0.5 in the odd site 1, the connectivity table $C^d$ and the new $M^d = A(0)(-C^d)^{-1}$ table at the mid-gap of parent change to:

| $C^d$ | 1 | 3 | 5 | 7 | 9 | 2 | 4 | 6 | 8 | 10 |
|---|---|---|---|---|---|---|---|---|---|---|
| 1 | -0.5 | 0 | 0 | 0 | 0 | -1 | 0 | 0 | 0 | -1 |
| 3 | 0 | 0 | 0 | 0 | 0 | -1 | -1 | 0 | 0 | 0 |
| 5 | 0 | 0 | 0 | 0 | 0 | 0 | -1 | -1 | 0 | -1 |
| 7 | 0 | 0 | 0 | 0 | 0 | 0 | 0 | -1 | -1 | 0 |
| 9 | 0 | 0 | 0 | 0 | 0 | 0 | 0 | 0 | -1 | -1 |
| 2 | -1 | -1 | 0 | 0 | 0 | 0 | 0 | 0 | 0 | 0 |
| 4 | 0 | -1 | -1 | 0 | 0 | 0 | 0 | 0 | 0 | 0 |
| 6 | 0 | 0 | -1 | -1 | 0 | 0 | 0 | 0 | 0 | 0 |
| 8 | 0 | 0 | 0 | -1 | -1 | 0 | 0 | 0 | 0 | 0 |
| 10 | -1 | 0 | -1 | 0 | -1 | 0 | 0 | 0 | 0 | 0 |

| $M^d$ | 1 | 3 | 5 | 7 | 9 | 2 | 4 | 6 | 8 | 10 |
|---|---|---|---|---|---|---|---|---|---|---|
| 1 | 0 | 0 | 0 | 0 | 0 | -2 | 2 | -1 | 1 | -1 |
| 3 | 0 | 0 | 0 | 0 | 0 | -1 | -2 | 1 | -1 | 1 |
| 5 | 0 | 0 | 0 | 0 | 0 | 1 | -1 | -1 | 1 | -1 |
| 7 | 0 | 0 | 0 | 0 | 0 | -1 | 1 | -2 | -1 | 1 |
| 9 | 0 | 0 | 0 | 0 | 0 | 1 | -1 | 2 | -2 | -1 |
| 2 | -2 | -1 | 1 | -1 | 1 | 2/3 | -2/3 | 1/3 | -1/3 | 1/3 |
| 4 | 2 | -2 | -1 | 1 | -1 | -2/3 | 2/3 | -1/3 | 1/3 | -1/3 |
| 6 | -1 | 1 | -1 | -2 | 2 | 1/3 | -1/3 | 1/6 | -1/6 | 1/6 |
| 8 | 1 | -1 | 1 | -1 | -2 | -1/3 | 1/3 | -1/6 | 1/6 | -1/6 |
| 10 | -1 | 1 | -1 | 1 | -1 | 1/3 | -1/3 | 1/6 | -1/6 | 1/6 |

Since $\tau_{ij}(0) = (M_{ij}(0))^2$, for $i$ and $j$ odd, both parent and daughter have low conductances (rule 1), if $i$ is odd and $j$ is even, or vice versa both parent and daughter have high conductances (rule 2), if $i,j$ are both even, the parent has a low conductance and the daughter a high conductance (rule 3) and if either $i$ or $j$ are odd, then $\tau_{ij}^d(0) = \tau_{ij}^p(0)$. Examples of other molecules are presented in the SI.

**(c)  DFT-based computational methods**

The optimized geometry and ground state Hamiltonian and overlap matrix elements of each structure was self-consistently obtained using the SIESTA[33] implementation of density functional theory (DFT). SIESTA employs norm-conserving pseudo-potentials to account for the core electrons and linear combinations of atomic orbitals to construct the valence states. The generalized gradient approximation (GGA) of the exchange and correlation functional is used with the Perdew-Burke-Ernzerhof parameterization (PBE)[34] a double-ζ polarized (DZP) basis set, a real-space grid defined with an equivalent energy cut-off of 250 Ry. The geometry optimization for each structure is performed to the forces smaller than 40 meV/Ang. Figure 3 shows examples of geometry-optimized structures used to obtain the DFT results in figures 4 and 5. The mean-field Hamiltonian obtained from the converged DFT calculation or a simple tight-binding Hamiltonian was combined with our Gollum[32] quantum transport code to calculate the phase-coherent, elastic scattering properties of the each system consist of left (source) and right (drain) leads and the scattering region. The transmission coefficient $T(E)$ for electrons of energy $E$ (passing from the source to the drain) is calculated via the relation $T(E) = Trace(\Gamma_R(E)G^R(E)\Gamma_L(E)G^{R\dagger}(E))$. In this expression, $\Gamma_{L,R}(E) = i\left(\sum_{L,R}(E) - \sum_{L,R}^{\dagger}(E)\right)$ describe the level broadening due to the coupling between left (L) and right (R) electrodes and the central scattering region, $\sum_{L,R}(E)$ are the retarded self-energies associated with this coupling and $G^R = (ES - H - \sum_L - \sum_R)^{-1}$ is the retarded Green's function, where $H$ is the Hamiltonian and $S$ is overlap matrix. Using obtained transmission coefficient ($T(E)$), the conductance could be calculated by Landauer formula ($G = G_0 \int dE\, T(E)(-\partial f/\partial E)$) where $G_0 = 2e^2/h$ is the conductance quantum, $f(E) = (1 + \exp((E - E_F)/k_B T))^{-1}$ is the Fermi-Dirac distribution function, $T$ is the temperature and $k_B$= 8.6x10$^{-5}$ eV/K is Boltzmann's constant.

**Author Information**

Corresponding Authors

*s.sangtarash@lancaster.ac.uk*

*c.lambert@lancaster.ac.uk*

**Acknowledgments**

This work was supported by the European Commission (EC) FP7 ITN "MOLESCO" (project no. 606728) and UK EPSRC (grant nos. EP/M014452/1 and EP/N017188/1).

*Supporting information*

# Exploring quantum interference in heteroatom-substituted graphene-like molecule s

Sara Sangtarash[*], Hatef Sadeghi, and Colin J. Lambert[*]

Quantum Technology Centre, Physics Department, Lancaster University, Lancaster LA1 4YB, UK

**s.sangtarash@lancaster.ac.uk; *c.lambert@lancaster.ac.uk*

1- For a naphthalene parental core, this shows the Hamiltonian $H^p$ and mid-gap (E=0) M-table $M^p$ of the parent, along with the Hamiltonian $H^d$ and E=0 M-table $M^d$ of the corresponding daughter.

| $H^p$ | 1 | 3 | 5 | 7 | 9 | 2 | 4 | 6 | 8 | 10 |
|---|---|---|---|---|---|---|---|---|---|---|
| 1 | 0 | 0 | 0 | 0 | 0 | -1 | 0 | 0 | 0 | -1 |
| 3 | 0 | 0 | 0 | 0 | 0 | -1 | -1 | 0 | 0 | 0 |
| 5 | 0 | 0 | 0 | 0 | 0 | 0 | -1 | -1 | 0 | -1 |
| 7 | 0 | 0 | 0 | 0 | 0 | 0 | 0 | -1 | -1 | 0 |
| 9 | 0 | 0 | 0 | 0 | 0 | 0 | 0 | 0 | -1 | -1 |
| 2 | -1 | -1 | 0 | 0 | 0 | 0 | 0 | 0 | 0 | 0 |
| 4 | 0 | -1 | -1 | 0 | 0 | 0 | 0 | 0 | 0 | 0 |
| 6 | 0 | 0 | -1 | -1 | 0 | 0 | 0 | 0 | 0 | 0 |
| 8 | 0 | 0 | 0 | -1 | -1 | 0 | 0 | 0 | 0 | 0 |
| 10 | -1 | 0 | -1 | 0 | -1 | 0 | 0 | 0 | 0 | 0 |

| $M^p$ | 1 | 3 | 5 | 7 | 9 | 2 | 4 | 6 | 8 | 10 |
|---|---|---|---|---|---|---|---|---|---|---|
| 1 | 0 | 0 | 0 | 0 | 0 | -2 | 2 | -1 | 1 | -1 |
| 3 | 0 | 0 | 0 | 0 | 0 | -1 | -2 | 1 | -1 | 1 |
| 5 | 0 | 0 | 0 | 0 | 0 | 1 | -1 | -1 | 1 | -1 |
| 7 | 0 | 0 | 0 | 0 | 0 | -1 | 1 | -2 | -1 | 1 |
| 9 | 0 | 0 | 0 | 0 | 0 | 1 | -1 | 2 | -2 | -1 |
| 2 | -2 | -1 | 1 | -1 | 1 | 0 | 0 | 0 | 0 | 0 |
| 4 | 2 | -2 | -1 | 1 | -1 | 0 | 0 | 0 | 0 | 0 |
| 6 | -1 | 1 | -1 | -2 | 2 | 0 | 0 | 0 | 0 | 0 |
| 8 | 1 | -1 | 1 | -1 | -2 | 0 | 0 | 0 | 0 | 0 |
| 10 | -1 | 1 | -1 | 1 | -1 | 0 | 0 | 0 | 0 | 0 |

| $H^d$ | 1 | 3 | 5 | 7 | 9 | 2 | 4 | 6 | 8 | 10 |
|---|---|---|---|---|---|---|---|---|---|---|
| 1 | -0.5 | 0 | 0 | 0 | 0 | -1 | 0 | 0 | 0 | -1 |
| 3 | 0 | 0 | 0 | 0 | 0 | -1 | -1 | 0 | 0 | 0 |
| 5 | 0 | 0 | 0 | 0 | 0 | 0 | -1 | -1 | 0 | -1 |
| 7 | 0 | 0 | 0 | 0 | 0 | 0 | 0 | -1 | -1 | 0 |
| 9 | 0 | 0 | 0 | 0 | 0 | 0 | 0 | 0 | -1 | -1 |

| | 2 | -1 | -1 | 0 | 0 | 0 | 0 | 0 | 0 | 0 | 0 |
|---|---|---|---|---|---|---|---|---|---|---|---|
| | 4 | 0 | -1 | -1 | 0 | 0 | 0 | 0 | 0 | 0 | 0 |
| | 6 | 0 | 0 | -1 | -1 | 0 | 0 | 0 | 0 | 0 | 0 |
| | 8 | 0 | 0 | 0 | -1 | -1 | 0 | 0 | 0 | 0 | 0 |
| | 10 | -1 | 0 | -1 | 0 | -1 | 0 | 0 | 0 | 0 | 0 |

| $M^d$ | 1 | 3 | 5 | 7 | 9 | 2 | 4 | 6 | 8 | 10 |
|---|---|---|---|---|---|---|---|---|---|---|
| 1 | 0 | 0 | 0 | 0 | 0 | -2 | 2 | -1 | 1 | -1 |
| 3 | 0 | 0 | 0 | 0 | 0 | -1 | -2 | 1 | -1 | 1 |
| 5 | 0 | 0 | 0 | 0 | 0 | 1 | -1 | -1 | 1 | -1 |
| 7 | 0 | 0 | 0 | 0 | 0 | -1 | 1 | -2 | -1 | 1 |
| 9 | 0 | 0 | 0 | 0 | 0 | 1 | -1 | 2 | -2 | -1 |
| 2 | -2 | -1 | 1 | -1 | 1 | 0.67 | -0.67 | 0.33 | -0.33 | 0.33 |
| 4 | 2 | -2 | -1 | 1 | -1 | -0.67 | 0.67 | -0.33 | 0.33 | -0.33 |
| 6 | -1 | 1 | -1 | -2 | 2 | 0.33 | -0.33 | 0.17 | -0.17 | 0.17 |
| 8 | 1 | -1 | 1 | -1 | -2 | -0.33 | 0.33 | -0.17 | 0.17 | -0.17 |
| 10 | -1 | 1 | -1 | 1 | -1 | 0.33 | -0.33 | 0.17 | -0.17 | 0.17 |

2- For a anthracene parental core, this shows the Hamiltonian $H^p$ and mid-gap (E=0) M-table $M^p$ of the parent, along with the Hamiltonian $H^d$ and E=0 M-table $M^d$ of the corresponding daughter.

| $H^p$ | 1 | 3 | 5 | 7 | 9 | 11 | 13 | 2 | 4 | 6 | 8 | 10 | 12 | 14 |
|---|---|---|---|---|---|---|---|---|---|---|---|---|---|---|
| 1 | 0 | 0 | 0 | 0 | 0 | 0 | 0 | -1 | 0 | 0 | 0 | 0 | 0 | -1 |
| 3 | 0 | 0 | 0 | 0 | 0 | 0 | 0 | -1 | -1 | 0 | 0 | 0 | 0 | 0 |
| 5 | 0 | 0 | 0 | 0 | 0 | 0 | 0 | 0 | -1 | -1 | 0 | 0 | 0 | 0 |
| 7 | 0 | 0 | 0 | 0 | 0 | 0 | 0 | -1 | 0 | -1 | -1 | 0 | 0 | 0 |
| 9 | 0 | 0 | 0 | 0 | 0 | 0 | 0 | 0 | 0 | 0 | -1 | -1 | 0 | -1 |
| 11 | 0 | 0 | 0 | 0 | 0 | 0 | 0 | 0 | 0 | 0 | 0 | -1 | -1 | 0 |
| 13 | 0 | 0 | 0 | 0 | 0 | 0 | 0 | 0 | 0 | 0 | 0 | 0 | -1 | -1 |
| 2 | -1 | -1 | 0 | -1 | 0 | 0 | 0 | 0 | 0 | 0 | 0 | 0 | 0 | 0 |
| 4 | 0 | -1 | -1 | 0 | 0 | 0 | 0 | 0 | 0 | 0 | 0 | 0 | 0 | 0 |
| 6 | 0 | 0 | -1 | -1 | 0 | 0 | 0 | 0 | 0 | 0 | 0 | 0 | 0 | 0 |
| 8 | 0 | 0 | 0 | -1 | -1 | 0 | 0 | 0 | 0 | 0 | 0 | 0 | 0 | 0 |
| 10 | 0 | 0 | 0 | 0 | -1 | -1 | 0 | 0 | 0 | 0 | 0 | 0 | 0 | 0 |
| 12 | 0 | 0 | 0 | 0 | 0 | -1 | -1 | 0 | 0 | 0 | 0 | 0 | 0 | 0 |
| 14 | -1 | 0 | 0 | 0 | -1 | 0 | -1 | 0 | 0 | 0 | 0 | 0 | 0 | 0 |

| $M^p$ | 1 | 3 | 5 | 7 | 9 | 11 | 13 | 2 | 4 | 6 | 8 | 10 | 12 | 14 |
|---|---|---|---|---|---|---|---|---|---|---|---|---|---|---|
| 1 | 0 | 0 | 0 | 0 | 0 | 0 | 0 | -2 | 2 | -2 | 4 | -2 | 2 | -2 |
| 3 | 0 | 0 | 0 | 0 | 0 | 0 | 0 | -1 | -3 | 3 | -2 | 1 | -1 | 1 |
| 5 | 0 | 0 | 0 | 0 | 0 | 0 | 0 | 1 | -1 | -3 | 2 | -1 | 1 | -1 |
| 7 | 0 | 0 | 0 | 0 | 0 | 0 | 0 | -1 | 1 | -1 | -2 | 1 | -1 | 1 |
| 9 | 0 | 0 | 0 | 0 | 0 | 0 | 0 | 1 | -1 | 1 | -2 | -1 | 1 | -1 |
| 11 | 0 | 0 | 0 | 0 | 0 | 0 | 0 | -1 | 1 | -1 | 2 | -3 | -1 | 1 |
| 13 | 0 | 0 | 0 | 0 | 0 | 0 | 0 | 1 | -1 | 1 | -2 | 3 | -3 | -1 |
| 2 | -2 | -1 | 1 | -1 | 1 | -1 | 1 | 0 | 0 | 0 | 0 | 0 | 0 | 0 |
| 4 | 2 | -3 | -1 | 1 | -1 | 1 | -1 | 0 | 0 | 0 | 0 | 0 | 0 | 0 |
| 6 | -2 | 3 | -3 | -1 | 1 | -1 | 1 | 0 | 0 | 0 | 0 | 0 | 0 | 0 |
| 8 | 4 | -2 | 2 | -2 | -2 | 2 | -2 | 0 | 0 | 0 | 0 | 0 | 0 | 0 |
| 10 | -2 | 1 | -1 | 1 | -1 | -3 | 3 | 0 | 0 | 0 | 0 | 0 | 0 | 0 |
| 12 | 2 | -1 | 1 | -1 | 1 | -1 | -3 | 0 | 0 | 0 | 0 | 0 | 0 | 0 |
| 14 | -2 | 1 | -1 | 1 | -1 | 1 | -1 | 0 | 0 | 0 | 0 | 0 | 0 | 0 |

| $H^d$ | 1 | 3 | 5 | 7 | 9 | 11 | 13 | 2 | 4 | 6 | 8 | 10 | 12 | 14 |
|---|---|---|---|---|---|---|---|---|---|---|---|---|---|---|
| 1 | -0.5 | 0 | 0 | 0 | 0 | 0 | 0 | -1 | 0 | 0 | 0 | 0 | 0 | -1 |
| 3 | 0 | 0 | 0 | 0 | 0 | 0 | 0 | -1 | -1 | 0 | 0 | 0 | 0 | 0 |
| 5 | 0 | 0 | 0 | 0 | 0 | 0 | 0 | 0 | -1 | -1 | 0 | 0 | 0 | 0 |
| 7 | 0 | 0 | 0 | 0 | 0 | 0 | 0 | -1 | 0 | -1 | -1 | 0 | 0 | 0 |
| 9 | 0 | 0 | 0 | 0 | 0 | 0 | 0 | 0 | 0 | 0 | -1 | -1 | 0 | -1 |
| 11 | 0 | 0 | 0 | 0 | 0 | 0 | 0 | 0 | 0 | 0 | 0 | -1 | -1 | 0 |
| 13 | 0 | 0 | 0 | 0 | 0 | 0 | 0 | 0 | 0 | 0 | 0 | 0 | -1 | -1 |
| 2 | -1 | -1 | 0 | -1 | 0 | 0 | 0 | 0 | 0 | 0 | 0 | 0 | 0 | 0 |
| 4 | 0 | -1 | -1 | 0 | 0 | 0 | 0 | 0 | 0 | 0 | 0 | 0 | 0 | 0 |
| 6 | 0 | 0 | -1 | -1 | 0 | 0 | 0 | 0 | 0 | 0 | 0 | 0 | 0 | 0 |
| 8 | 0 | 0 | 0 | -1 | -1 | 0 | 0 | 0 | 0 | 0 | 0 | 0 | 0 | 0 |
| 10 | 0 | 0 | 0 | 0 | -1 | -1 | 0 | 0 | 0 | 0 | 0 | 0 | 0 | 0 |
| 12 | 0 | 0 | 0 | 0 | 0 | -1 | -1 | 0 | 0 | 0 | 0 | 0 | 0 | 0 |
| 14 | -1 | 0 | 0 | 0 | -1 | 0 | -1 | 0 | 0 | 0 | 0 | 0 | 0 | 0 |

| $M^d$ | 1 | 3 | 5 | 7 | 9 | 11 | 13 | 2 | 4 | 6 | 8 | 10 | 12 | 14 |
|---|---|---|---|---|---|---|---|---|---|---|---|---|---|---|
| 1 | 0 | 0 | 0 | 0 | 0 | 0 | 0 | -2 | 2 | -2 | 4 | -2 | 2 | -2 |
| 3 | 0 | 0 | 0 | 0 | 0 | 0 | 0 | -1 | -3 | 3 | -2 | 1 | -1 | 1 |
| 5 | 0 | 0 | 0 | 0 | 0 | 0 | 0 | 1 | -1 | -3 | 2 | -1 | 1 | -1 |
| 7 | 0 | 0 | 0 | 0 | 0 | 0 | 0 | -1 | 1 | -1 | -2 | 1 | -1 | 1 |
| 9 | 0 | 0 | 0 | 0 | 0 | 0 | 0 | 1 | -1 | 1 | -2 | -1 | 1 | -1 |
| 11 | 0 | 0 | 0 | 0 | 0 | 0 | 0 | -1 | 1 | -1 | 2 | -3 | -1 | 1 |
| 13 | 0 | 0 | 0 | 0 | 0 | 0 | 0 | 1 | -1 | 1 | -2 | 3 | -3 | -1 |
| 2 | -2 | -1 | 1 | -1 | 1 | -1 | 1 | 0.5 | -0.5 | 0.5 | -1 | 0.5 | -0.5 | 0.5 |
| 4 | 2 | -3 | -1 | 1 | -1 | 1 | -1 | -0.5 | 0.5 | -0.5 | 1 | -0.5 | 0.5 | -0.5 |
| 6 | -2 | 3 | -3 | -1 | 1 | -1 | 1 | 0.5 | -0.5 | 0.5 | -1 | 0.5 | -0.5 | 0.5 |
| 8 | 4 | -2 | 2 | -2 | -2 | 2 | -2 | -1 | 1 | -1 | 2 | -1 | 1 | -1 |
| 10 | -2 | 1 | -1 | 1 | -1 | -3 | 3 | 0.5 | -0.5 | 0.5 | -1 | 0.5 | -0.5 | 0.5 |
| 12 | 2 | -1 | 1 | -1 | 1 | -1 | -3 | -0.5 | 0.5 | -0.5 | 1 | -0.5 | 0.5 | -0.5 |
| 14 | -2 | 1 | -1 | 1 | -1 | 1 | -1 | 0.5 | -0.5 | 0.5 | -1 | 0.5 | -0.5 | 0.5 |

3- For a pyrene parental core, this shows the Hamiltonian $H^p$ and mid-gap (E=0) M-table $M^p$ of the parent, along with the Hamiltonian $H^d$ and E=0 M-table $M^d$ of the corresponding daughter.

| $H^p$ | 1 | 3 | 5 | 7 | 9 | 11 | 13 | 15 | 2 | 4 | 6 | 8 | 10 | 12 | 14 | 16 |
|---|---|---|---|---|---|---|---|---|---|---|---|---|---|---|---|---|
| 1 | 0 | 0 | 0 | 0 | 0 | 0 | 0 | 0 | -1 | 0 | 0 | 0 | 0 | 0 | -1 | 0 |
| 3 | 0 | 0 | 0 | 0 | 0 | 0 | 0 | 0 | -1 | -1 | 0 | 0 | 0 | 0 | 0 | -1 |
| 5 | 0 | 0 | 0 | 0 | 0 | 0 | 0 | 0 | 0 | -1 | -1 | 0 | 0 | 0 | 0 | 0 |
| 7 | 0 | 0 | 0 | 0 | 0 | 0 | 0 | 0 | 0 | 0 | -1 | -1 | 0 | 0 | 0 | -1 |
| 9 | 0 | 0 | 0 | 0 | 0 | 0 | 0 | 0 | 0 | 0 | 0 | -1 | -1 | 0 | 0 | 0 |
| 11 | 0 | 0 | 0 | 0 | 0 | 0 | 0 | 0 | 0 | 0 | 0 | 0 | -1 | -1 | 0 | 0 |
| 13 | 0 | 0 | 0 | 0 | 0 | 0 | 0 | 0 | 0 | 0 | 0 | 0 | 0 | -1 | -1 | 0 |
| 15 | 0 | 0 | 0 | 0 | 0 | 0 | 0 | 0 | 0 | 0 | 0 | 0 | -1 | 0 | -1 | -1 |
| 2 | -1 | -1 | 0 | 0 | 0 | 0 | 0 | 0 | 0 | 0 | 0 | 0 | 0 | 0 | 0 | 0 |
| 4 | 0 | -1 | -1 | 0 | 0 | 0 | 0 | 0 | 0 | 0 | 0 | 0 | 0 | 0 | 0 | 0 |
| 6 | 0 | 0 | -1 | -1 | 0 | 0 | 0 | 0 | 0 | 0 | 0 | 0 | 0 | 0 | 0 | 0 |
| 8 | 0 | 0 | 0 | -1 | -1 | 0 | 0 | 0 | 0 | 0 | 0 | 0 | 0 | 0 | 0 | 0 |
| 10 | 0 | 0 | 0 | 0 | -1 | -1 | 0 | -1 | 0 | 0 | 0 | 0 | 0 | 0 | 0 | 0 |
| 12 | 0 | 0 | 0 | 0 | 0 | -1 | -1 | 0 | 0 | 0 | 0 | 0 | 0 | 0 | 0 | 0 |

| 14 | -1 | 0 | 0 | 0 | 0 | 0 | -1 | -1 | 0 | 0 | 0 | 0 | 0 | 0 | 0 | 0 |
|---|---|---|---|---|---|---|---|---|---|---|---|---|---|---|---|---|
| 16 | 0 | -1 | 0 | -1 | 0 | 0 | 0 | -1 | 0 | 0 | 0 | 0 | 0 | 0 | 0 | 0 |

| $M^p$ | 1 | 3 | 5 | 7 | 9 | 11 | 13 | 15 | 2 | 4 | 6 | 8 | 10 | 12 | 14 | 16 |
|---|---|---|---|---|---|---|---|---|---|---|---|---|---|---|---|---|
| 1 | 0 | 0 | 0 | 0 | 0 | 0 | 0 | 0 | -5 | 3 | -3 | 1 | -1 | 1 | -1 | 2 |
| 3 | 0 | 0 | 0 | 0 | 0 | 0 | 0 | 0 | -1 | -3 | 3 | -1 | 1 | -1 | 1 | -2 |
| 5 | 0 | 0 | 0 | 0 | 0 | 0 | 0 | 0 | 1 | -3 | -3 | 1 | -1 | 1 | -1 | 2 |
| 7 | 0 | 0 | 0 | 0 | 0 | 0 | 0 | 0 | -1 | 3 | -3 | -1 | 1 | -1 | 1 | -2 |
| 9 | 0 | 0 | 0 | 0 | 0 | 0 | 0 | 0 | 1 | -3 | 3 | -5 | -1 | 1 | -1 | 2 |
| 11 | 0 | 0 | 0 | 0 | 0 | 0 | 0 | 0 | -3 | 3 | -3 | 3 | -3 | -3 | 3 | 0 |
| 13 | 0 | 0 | 0 | 0 | 0 | 0 | 0 | 0 | 3 | -3 | 3 | -3 | 3 | -3 | -3 | 0 |
| 15 | 0 | 0 | 0 | 0 | 0 | 0 | 0 | 0 | 2 | 0 | 0 | 2 | -2 | 2 | -2 | -2 |
| 2 | -5 | -1 | 1 | -1 | 1 | -3 | 3 | 2 | 0 | 0 | 0 | 0 | 0 | 0 | 0 | 0 |
| 4 | 3 | -3 | -3 | 3 | -3 | 3 | -3 | 0 | 0 | 0 | 0 | 0 | 0 | 0 | 0 | 0 |
| 6 | -3 | 3 | -3 | -3 | 3 | -3 | 3 | 0 | 0 | 0 | 0 | 0 | 0 | 0 | 0 | 0 |
| 8 | 1 | -1 | 1 | -1 | -5 | 3 | -3 | 2 | 0 | 0 | 0 | 0 | 0 | 0 | 0 | 0 |
| 10 | -1 | 1 | -1 | 1 | -1 | -3 | 3 | -2 | 0 | 0 | 0 | 0 | 0 | 0 | 0 | 0 |
| 12 | 1 | -1 | 1 | -1 | 1 | -3 | -3 | 2 | 0 | 0 | 0 | 0 | 0 | 0 | 0 | 0 |
| 14 | -1 | 1 | -1 | 1 | -1 | 3 | -3 | -2 | 0 | 0 | 0 | 0 | 0 | 0 | 0 | 0 |
| 16 | 2 | -2 | 2 | -2 | 2 | 0 | 0 | -2 | 0 | 0 | 0 | 0 | 0 | 0 | 0 | 0 |

| $H^d$ | 1 | 3 | 5 | 7 | 9 | 11 | 13 | 15 | 2 | 4 | 6 | 8 | 10 | 12 | 14 | 16 |
|---|---|---|---|---|---|---|---|---|---|---|---|---|---|---|---|---|
| 1 | -0.5 | 0 | 0 | 0 | 0 | 0 | 0 | 0 | -1 | 0 | 0 | 0 | 0 | 0 | -1 | 0 |
| 3 | 0 | 0 | 0 | 0 | 0 | 0 | 0 | 0 | -1 | -1 | 0 | 0 | 0 | 0 | 0 | -1 |
| 5 | 0 | 0 | 0 | 0 | 0 | 0 | 0 | 0 | 0 | -1 | -1 | 0 | 0 | 0 | 0 | 0 |
| 7 | 0 | 0 | 0 | 0 | 0 | 0 | 0 | 0 | 0 | 0 | -1 | -1 | 0 | 0 | 0 | -1 |
| 9 | 0 | 0 | 0 | 0 | 0 | 0 | 0 | 0 | 0 | 0 | 0 | -1 | -1 | 0 | 0 | 0 |
| 11 | 0 | 0 | 0 | 0 | 0 | 0 | 0 | 0 | 0 | 0 | 0 | 0 | -1 | -1 | 0 | 0 |
| 13 | 0 | 0 | 0 | 0 | 0 | 0 | 0 | 0 | 0 | 0 | 0 | 0 | 0 | -1 | -1 | 0 |
| 15 | 0 | 0 | 0 | 0 | 0 | 0 | 0 | 0 | 0 | 0 | 0 | 0 | -1 | 0 | -1 | -1 |
| 2 | -1 | -1 | 0 | 0 | 0 | 0 | 0 | 0 | 0 | 0 | 0 | 0 | 0 | 0 | 0 | 0 |
| 4 | 0 | -1 | -1 | 0 | 0 | 0 | 0 | 0 | 0 | 0 | 0 | 0 | 0 | 0 | 0 | 0 |
| 6 | 0 | 0 | -1 | -1 | 0 | 0 | 0 | 0 | 0 | 0 | 0 | 0 | 0 | 0 | 0 | 0 |
| 8 | 0 | 0 | 0 | -1 | -1 | 0 | 0 | 0 | 0 | 0 | 0 | 0 | 0 | 0 | 0 | 0 |
| 10 | 0 | 0 | 0 | 0 | -1 | -1 | 0 | -1 | 0 | 0 | 0 | 0 | 0 | 0 | 0 | 0 |
| 12 | 0 | 0 | 0 | 0 | 0 | -1 | -1 | 0 | 0 | 0 | 0 | 0 | 0 | 0 | 0 | 0 |
| 14 | -1 | 0 | 0 | 0 | 0 | 0 | -1 | -1 | 0 | 0 | 0 | 0 | 0 | 0 | 0 | 0 |
| 16 | 0 | -1 | 0 | -1 | 0 | 0 | 0 | -1 | 0 | 0 | 0 | 0 | 0 | 0 | 0 | 0 |

| $M^d$ | 1 | 3 | 5 | 7 | 9 | 11 | 13 | 15 | 2 | 4 | 6 | 8 | 10 | 12 | 14 | 16 |
|---|---|---|---|---|---|---|---|---|---|---|---|---|---|---|---|---|
| 1 | 0 | 0 | 0 | 0 | 0 | 0 | 0 | 0 | -5 | 3 | -3 | 1 | -1 | 1 | -1 | 2 |
| 3 | 0 | 0 | 0 | 0 | 0 | 0 | 0 | 0 | -1 | -3 | 3 | -1 | 1 | -1 | 1 | -2 |
| 5 | 0 | 0 | 0 | 0 | 0 | 0 | 0 | 0 | 1 | -3 | -3 | 1 | -1 | 1 | -1 | 2 |
| 7 | 0 | 0 | 0 | 0 | 0 | 0 | 0 | 0 | -1 | 3 | -3 | -1 | 1 | -1 | 1 | -2 |
| 9 | 0 | 0 | 0 | 0 | 0 | 0 | 0 | 0 | 1 | -3 | 3 | -5 | -1 | 1 | -1 | 2 |
| 11 | 0 | 0 | 0 | 0 | 0 | 0 | 0 | 0 | -3 | 3 | -3 | 3 | -3 | -3 | 3 | 0 |
| 13 | 0 | 0 | 0 | 0 | 0 | 0 | 0 | 0 | 3 | -3 | 3 | -3 | 3 | -3 | -3 | 0 |
| 15 | 0 | 0 | 0 | 0 | 0 | 0 | 0 | 0 | 2 | 0 | 0 | 2 | -2 | 2 | -2 | -2 |
| 2 | -5 | -1 | 1 | -1 | 1 | -3 | 3 | 2 | 2.08 | -1.25 | 1.25 | -0.42 | 0.42 | -0.42 | 0.42 | -0.83 |
| 4 | 3 | -3 | -3 | 3 | -3 | 3 | -3 | 0 | -1.25 | 0.75 | -0.75 | 0.25 | -0.25 | 0.25 | -0.25 | 0.50 |
| 6 | -3 | 3 | -3 | -3 | 3 | -3 | 3 | 0 | 1.25 | -0.75 | 0.75 | -0.25 | 0.25 | -0.25 | 0.25 | -0.50 |
| 8 | 1 | -1 | 1 | -1 | -5 | 3 | -3 | 2 | -0.42 | 0.25 | -0.25 | 0.08 | -0.08 | 0.08 | -0.08 | 0.17 |
| 10 | -1 | 1 | -1 | 1 | -1 | -3 | 3 | -2 | 0.42 | -0.25 | 0.25 | -0.08 | 0.08 | -0.08 | 0.08 | -0.17 |
| 12 | 1 | -1 | 1 | -1 | 1 | -3 | -3 | 2 | -0.42 | 0.25 | -0.25 | 0.08 | -0.08 | 0.08 | -0.08 | 0.17 |
| 14 | -1 | 1 | -1 | 1 | -1 | 3 | -3 | -2 | 0.42 | -0.25 | 0.25 | -0.08 | 0.08 | -0.08 | 0.08 | -0.17 |
| 16 | 2 | -2 | 2 | -2 | 2 | 0 | 0 | -2 | -0.83 | 0.50 | -0.50 | 0.17 | -0.17 | 0.17 | -0.17 | 0.33 |

4- For a anthanthrene parental core, this shows the Hamiltonian $H^p$ and mid-gap (E=0) M-table $M^p$ of the parent, along with the Hamiltonian $H^d$ and E=0 M-table $M^d$ of the corresponding daughter.

| $H^p$ | 1 | 3 | 5 | 7 | 9 | 11 | 13 | 15 | 17 | 19 | 21 | 2 | 4 | 6 | 8 | 10 | 12 | 14 | 16 | 18 | 20 | 22 |
|---|---|---|---|---|---|---|---|---|---|---|---|---|---|---|---|---|---|---|---|---|---|---|
| 1 | 0 | 0 | 0 | 0 | 0 | 0 | 0 | 0 | 0 | 0 | 0 | -1 | 0 | 0 | 0 | 0 | 0 | 0 | 0 | 0 | 0 | -1 |
| 3 | 0 | 0 | 0 | 0 | 0 | 0 | 0 | 0 | 0 | 0 | 0 | -1 | -1 | 0 | 0 | 0 | 0 | 0 | 0 | 0 | 0 | 0 |
| 5 | 0 | 0 | 0 | 0 | 0 | 0 | 0 | 0 | 0 | 0 | 0 | 0 | -1 | -1 | 0 | 0 | 0 | 0 | 0 | 0 | 0 | 0 |
| 7 | 0 | 0 | 0 | 0 | 0 | 0 | 0 | 0 | 0 | 0 | 0 | 0 | 0 | -1 | -1 | 0 | 0 | 0 | 0 | 0 | 0 | 0 |
| 9 | 0 | 0 | 0 | 0 | 0 | 0 | 0 | 0 | 0 | 0 | 0 | 0 | 0 | 0 | -1 | -1 | 0 | 0 | 0 | 0 | 0 | 0 |
| 11 | 0 | 0 | 0 | 0 | 0 | 0 | 0 | 0 | 0 | 0 | 0 | 0 | 0 | 0 | 0 | -1 | -1 | 0 | 0 | 0 | -1 | 0 |
| 13 | 0 | 0 | 0 | 0 | 0 | 0 | 0 | 0 | 0 | 0 | 0 | 0 | 0 | 0 | 0 | 0 | -1 | -1 | 0 | -1 | 0 | 0 |
| 15 | 0 | 0 | 0 | 0 | 0 | 0 | 0 | 0 | 0 | 0 | 0 | 0 | 0 | 0 | 0 | 0 | 0 | -1 | -1 | 0 | 0 | 0 |
| 17 | 0 | 0 | 0 | 0 | 0 | 0 | 0 | 0 | 0 | 0 | 0 | 0 | 0 | 0 | 0 | 0 | 0 | 0 | -1 | -1 | 0 | -1 |
| 19 | 0 | 0 | 0 | 0 | 0 | 0 | 0 | 0 | 0 | 0 | 0 | -1 | 0 | 0 | 0 | 0 | 0 | 0 | 0 | -1 | -1 | 0 |
| 21 | 0 | 0 | 0 | 0 | 0 | 0 | 0 | 0 | 0 | 0 | 0 | 0 | -1 | 0 | -1 | 0 | 0 | 0 | 0 | 0 | -1 | 0 |
| 2 | -1 | -1 | 0 | 0 | 0 | 0 | 0 | 0 | 0 | -1 | 0 | 0 | 0 | 0 | 0 | 0 | 0 | 0 | 0 | 0 | 0 | 0 |
| 4 | 0 | -1 | -1 | 0 | 0 | 0 | 0 | 0 | 0 | 0 | -1 | 0 | 0 | 0 | 0 | 0 | 0 | 0 | 0 | 0 | 0 | 0 |
| 6 | 0 | 0 | -1 | -1 | 0 | 0 | 0 | 0 | 0 | 0 | 0 | 0 | 0 | 0 | 0 | 0 | 0 | 0 | 0 | 0 | 0 | 0 |
| 8 | 0 | 0 | 0 | -1 | -1 | 0 | 0 | 0 | 0 | 0 | -1 | 0 | 0 | 0 | 0 | 0 | 0 | 0 | 0 | 0 | 0 | 0 |
| 10 | 0 | 0 | 0 | 0 | -1 | -1 | 0 | 0 | 0 | 0 | 0 | 0 | 0 | 0 | 0 | 0 | 0 | 0 | 0 | 0 | 0 | 0 |
| 12 | 0 | 0 | 0 | 0 | 0 | -1 | -1 | 0 | 0 | 0 | 0 | 0 | 0 | 0 | 0 | 0 | 0 | 0 | 0 | 0 | 0 | 0 |
| 14 | 0 | 0 | 0 | 0 | 0 | 0 | -1 | -1 | 0 | 0 | 0 | 0 | 0 | 0 | 0 | 0 | 0 | 0 | 0 | 0 | 0 | 0 |
| 16 | 0 | 0 | 0 | 0 | 0 | 0 | 0 | -1 | -1 | 0 | 0 | 0 | 0 | 0 | 0 | 0 | 0 | 0 | 0 | 0 | 0 | 0 |
| 18 | 0 | 0 | 0 | 0 | 0 | 0 | -1 | 0 | -1 | -1 | 0 | 0 | 0 | 0 | 0 | 0 | 0 | 0 | 0 | 0 | 0 | 0 |
| 20 | 0 | 0 | 0 | 0 | 0 | -1 | 0 | 0 | 0 | -1 | -1 | 0 | 0 | 0 | 0 | 0 | 0 | 0 | 0 | 0 | 0 | 0 |
| 22 | -1 | 0 | 0 | 0 | 0 | 0 | 0 | -1 | 0 | 0 | 0 | 0 | 0 | 0 | 0 | 0 | 0 | 0 | 0 | 0 | 0 | 0 |

| $M^p$ | 1 | 3 | 5 | 7 | 9 | 11 | 13 | 15 | 17 | 19 | 21 | 2 | 4 | 6 | 8 | 10 | 12 | 14 | 16 | 18 | 20 | 22 |
|---|---|---|---|---|---|---|---|---|---|---|---|---|---|---|---|---|---|---|---|---|---|---|
| 1 | 0 | 0 | 0 | 0 | 0 | 0 | 0 | 0 | 0 | 0 | 0 | -1 | 1 | -1 | 1 | -1 | 3 | -6 | 6 | 3 | -2 | -9 |
| 3 | 0 | 0 | 0 | 0 | 0 | 0 | 0 | 0 | 0 | 0 | 0 | -7 | -3 | 3 | -3 | 3 | -9 | 8 | -8 | 1 | 6 | 7 |
| 5 | 0 | 0 | 0 | 0 | 0 | 0 | 0 | 0 | 0 | 0 | 0 | 4 | -4 | -6 | 6 | -6 | 8 | -6 | 6 | -2 | -2 | -4 |
| 7 | 0 | 0 | 0 | 0 | 0 | 0 | 0 | 0 | 0 | 0 | 0 | -4 | 4 | -4 | -6 | 6 | -8 | 6 | -6 | 2 | 2 | 4 |
| 9 | 0 | 0 | 0 | 0 | 0 | 0 | 0 | 0 | 0 | 0 | 0 | 1 | -1 | 1 | -1 | -9 | 7 | -4 | 4 | -3 | 2 | -1 |
| 11 | 0 | 0 | 0 | 0 | 0 | 0 | 0 | 0 | 0 | 0 | 0 | -1 | 1 | -1 | 1 | -1 | -7 | 4 | -4 | 3 | -2 | 1 |
| 13 | 0 | 0 | 0 | 0 | 0 | 0 | 0 | 0 | 0 | 0 | 0 | 1 | -1 | 1 | -1 | 1 | -3 | -4 | 4 | -3 | 2 | -1 |
| 15 | 0 | 0 | 0 | 0 | 0 | 0 | 0 | 0 | 0 | 0 | 0 | -1 | 1 | -1 | 1 | -1 | 3 | -6 | -4 | 3 | -2 | 1 |
| 17 | 0 | 0 | 0 | 0 | 0 | 0 | 0 | 0 | 0 | 0 | 0 | 1 | -1 | 1 | -1 | 1 | -3 | 6 | -6 | -3 | 2 | -1 |
| 19 | 0 | 0 | 0 | 0 | 0 | 0 | 0 | 0 | 0 | 0 | 0 | -2 | 2 | -2 | 2 | -2 | 6 | -2 | 2 | -4 | -4 | 2 |
| 21 | 0 | 0 | 0 | 0 | 0 | 0 | 0 | 0 | 0 | 0 | 0 | 3 | -3 | 3 | -3 | 3 | 1 | -2 | 2 | 1 | -4 | -3 |
| 2 | -1 | -7 | 4 | -4 | 1 | -1 | 1 | -1 | 1 | -2 | 3 | 0 | 0 | 0 | 0 | 0 | 0 | 0 | 0 | 0 | 0 | 0 |
| 4 | 1 | -3 | -4 | 4 | -1 | 1 | -1 | 1 | -1 | 2 | -3 | 0 | 0 | 0 | 0 | 0 | 0 | 0 | 0 | 0 | 0 | 0 |
| 6 | -1 | 3 | -6 | -4 | 1 | -1 | 1 | -1 | 1 | -2 | 3 | 0 | 0 | 0 | 0 | 0 | 0 | 0 | 0 | 0 | 0 | 0 |
| 8 | 1 | -3 | 6 | -6 | -1 | 1 | -1 | 1 | -1 | 2 | -3 | 0 | 0 | 0 | 0 | 0 | 0 | 0 | 0 | 0 | 0 | 0 |
| 10 | -1 | 3 | -6 | 6 | -9 | -1 | 1 | -1 | 1 | -2 | 3 | 0 | 0 | 0 | 0 | 0 | 0 | 0 | 0 | 0 | 0 | 0 |
| 12 | 3 | -9 | 8 | -8 | 7 | -7 | -3 | 3 | -3 | 6 | 1 | 0 | 0 | 0 | 0 | 0 | 0 | 0 | 0 | 0 | 0 | 0 |
| 14 | -6 | 8 | -6 | 6 | -4 | 4 | -4 | -6 | 6 | -2 | -2 | 0 | 0 | 0 | 0 | 0 | 0 | 0 | 0 | 0 | 0 | 0 |
| 16 | 6 | -8 | 6 | -6 | 4 | -4 | 4 | -4 | -6 | 2 | 2 | 0 | 0 | 0 | 0 | 0 | 0 | 0 | 0 | 0 | 0 | 0 |
| 18 | 3 | 1 | -2 | 2 | -3 | 3 | -3 | 3 | -3 | -4 | 1 | 0 | 0 | 0 | 0 | 0 | 0 | 0 | 0 | 0 | 0 | 0 |
| 20 | -2 | 6 | -2 | 2 | 2 | -2 | 2 | -2 | 2 | -4 | -4 | 0 | 0 | 0 | 0 | 0 | 0 | 0 | 0 | 0 | 0 | 0 |
| 22 | -9 | 7 | -4 | 4 | -1 | 1 | -1 | 1 | -1 | 2 | -3 | 0 | 0 | 0 | 0 | 0 | 0 | 0 | 0 | 0 | 0 | 0 |

| $H^d$ | 1 | 3 | 5 | 7 | 9 | 11 | 13 | 15 | 17 | 19 | 21 | 2 | 4 | 6 | 8 | 10 | 12 | 14 | 16 | 18 | 20 | 22 |
|---|---|---|---|---|---|---|---|---|---|---|---|---|---|---|---|---|---|---|---|---|---|---|
| 1 | -0.5 | 0 | 0 | 0 | 0 | 0 | 0 | 0 | 0 | 0 | 0 | -1 | 0 | 0 | 0 | 0 | 0 | 0 | 0 | 0 | 0 | -1 |
| 3 | 0 | 0 | 0 | 0 | 0 | 0 | 0 | 0 | 0 | 0 | 0 | -1 | -1 | 0 | 0 | 0 | 0 | 0 | 0 | 0 | 0 | 0 |
| 5 | 0 | 0 | 0 | 0 | 0 | 0 | 0 | 0 | 0 | 0 | 0 | 0 | -1 | -1 | 0 | 0 | 0 | 0 | 0 | 0 | 0 | 0 |
| 7 | 0 | 0 | 0 | 0 | 0 | 0 | 0 | 0 | 0 | 0 | 0 | 0 | 0 | -1 | -1 | 0 | 0 | 0 | 0 | 0 | 0 | 0 |
| 9 | 0 | 0 | 0 | 0 | 0 | 0 | 0 | 0 | 0 | 0 | 0 | 0 | 0 | 0 | -1 | -1 | 0 | 0 | 0 | 0 | 0 | 0 |
| 11 | 0 | 0 | 0 | 0 | 0 | 0 | 0 | 0 | 0 | 0 | 0 | 0 | 0 | 0 | 0 | -1 | -1 | 0 | 0 | 0 | -1 | 0 |
| 13 | 0 | 0 | 0 | 0 | 0 | 0 | 0 | 0 | 0 | 0 | 0 | 0 | 0 | 0 | 0 | 0 | -1 | -1 | 0 | -1 | 0 | 0 |
| 15 | 0 | 0 | 0 | 0 | 0 | 0 | 0 | 0 | 0 | 0 | 0 | 0 | 0 | 0 | 0 | 0 | 0 | -1 | -1 | 0 | 0 | 0 |
| 17 | 0 | 0 | 0 | 0 | 0 | 0 | 0 | 0 | 0 | 0 | 0 | 0 | 0 | 0 | 0 | 0 | 0 | 0 | -1 | -1 | 0 | -1 |
| 19 | 0 | 0 | 0 | 0 | 0 | 0 | 0 | 0 | 0 | 0 | 0 | -1 | 0 | 0 | 0 | 0 | 0 | 0 | 0 | -1 | -1 | 0 |
| 21 | 0 | 0 | 0 | 0 | 0 | 0 | 0 | 0 | 0 | 0 | 0 | 0 | -1 | 0 | -1 | 0 | 0 | 0 | 0 | 0 | -1 | 0 |
| 2 | -1 | -1 | 0 | 0 | 0 | 0 | 0 | 0 | -1 | 0 | 0 | 0 | 0 | 0 | 0 | 0 | 0 | 0 | 0 | 0 | 0 | 0 |
| 4 | 0 | -1 | -1 | 0 | 0 | 0 | 0 | 0 | 0 | 0 | -1 | 0 | 0 | 0 | 0 | 0 | 0 | 0 | 0 | 0 | 0 | 0 |

| | 1 | 3 | 5 | 7 | 9 | 11 | 13 | 15 | 17 | 19 | 21 | 2 | 4 | 6 | 8 | 10 | 12 | 14 | 16 | 18 | 20 | 22 |
|---|---|---|---|---|---|---|---|---|---|---|---|---|---|---|---|---|---|---|---|---|---|---|
| 6 | 0 | 0 | -1 | -1 | 0 | 0 | 0 | 0 | 0 | 0 | 0 | 0 | 0 | 0 | 0 | 0 | 0 | 0 | 0 | 0 | 0 | 0 |
| 8 | 0 | 0 | 0 | -1 | -1 | 0 | 0 | 0 | 0 | 0 | -1 | 0 | 0 | 0 | 0 | 0 | 0 | 0 | 0 | 0 | 0 | 0 |
| 10 | 0 | 0 | 0 | 0 | -1 | -1 | 0 | 0 | 0 | 0 | 0 | 0 | 0 | 0 | 0 | 0 | 0 | 0 | 0 | 0 | 0 | 0 |
| 12 | 0 | 0 | 0 | 0 | 0 | -1 | -1 | 0 | 0 | 0 | 0 | 0 | 0 | 0 | 0 | 0 | 0 | 0 | 0 | 0 | 0 | 0 |
| 14 | 0 | 0 | 0 | 0 | 0 | 0 | -1 | -1 | 0 | 0 | 0 | 0 | 0 | 0 | 0 | 0 | 0 | 0 | 0 | 0 | 0 | 0 |
| 16 | 0 | 0 | 0 | 0 | 0 | 0 | 0 | -1 | -1 | 0 | 0 | 0 | 0 | 0 | 0 | 0 | 0 | 0 | 0 | 0 | 0 | 0 |
| 18 | 0 | 0 | 0 | 0 | 0 | 0 | -1 | 0 | -1 | -1 | 0 | 0 | 0 | 0 | 0 | 0 | 0 | 0 | 0 | 0 | 0 | 0 |
| 20 | 0 | 0 | 0 | 0 | 0 | -1 | 0 | 0 | 0 | -1 | -1 | 0 | 0 | 0 | 0 | 0 | 0 | 0 | 0 | 0 | 0 | 0 |
| 22 | -1 | 0 | 0 | 0 | 0 | 0 | 0 | 0 | -1 | 0 | 0 | 0 | 0 | 0 | 0 | 0 | 0 | 0 | 0 | 0 | 0 | 0 |

| $M^d$ | 1 | 3 | 5 | 7 | 9 | 11 | 13 | 15 | 17 | 19 | 21 | 2 | 4 | 6 | 8 | 10 | 12 | 14 | 16 | 18 | 20 | 22 |
|---|---|---|---|---|---|---|---|---|---|---|---|---|---|---|---|---|---|---|---|---|---|---|
| 1 | 0 | 0 | 0 | 0 | 0 | 0 | 0 | 0 | 0 | 0 | 0 | -1 | 1 | -1 | 1 | -1 | 3 | -6 | 6 | 3 | -2 | -9 |
| 3 | 0 | 0 | 0 | 0 | 0 | 0 | 0 | 0 | 0 | 0 | 0 | -7 | -3 | 3 | -3 | 3 | -9 | 8 | -8 | 1 | 6 | 7 |
| 5 | 0 | 0 | 0 | 0 | 0 | 0 | 0 | 0 | 0 | 0 | 0 | 4 | -4 | -6 | 6 | -6 | 8 | -6 | 6 | -2 | -2 | -4 |
| 7 | 0 | 0 | 0 | 0 | 0 | 0 | 0 | 0 | 0 | 0 | 0 | -4 | 4 | -4 | -6 | 6 | -8 | 6 | -6 | 2 | 2 | 4 |
| 9 | 0 | 0 | 0 | 0 | 0 | 0 | 0 | 0 | 0 | 0 | 0 | 1 | -1 | 1 | -1 | -9 | 7 | -4 | 4 | -3 | 2 | -1 |
| 11 | 0 | 0 | 0 | 0 | 0 | 0 | 0 | 0 | 0 | 0 | 0 | -1 | 1 | -1 | 1 | -1 | -7 | 4 | -4 | 3 | -2 | 1 |
| 13 | 0 | 0 | 0 | 0 | 0 | 0 | 0 | 0 | 0 | 0 | 0 | 1 | -1 | 1 | -1 | 1 | -3 | -4 | 4 | -3 | 2 | -1 |
| 15 | 0 | 0 | 0 | 0 | 0 | 0 | 0 | 0 | 0 | 0 | 0 | -1 | 1 | -1 | 1 | -1 | 3 | -6 | -4 | 3 | -2 | 1 |
| 17 | 0 | 0 | 0 | 0 | 0 | 0 | 0 | 0 | 0 | 0 | 0 | 1 | -1 | 1 | -1 | 1 | -3 | 6 | -6 | -3 | 2 | -1 |
| 19 | 0 | 0 | 0 | 0 | 0 | 0 | 0 | 0 | 0 | 0 | 0 | -2 | 2 | -2 | 2 | -2 | 6 | -2 | 2 | -4 | -4 | 2 |
| 21 | 0 | 0 | 0 | 0 | 0 | 0 | 0 | 0 | 0 | 0 | 0 | 3 | -3 | 3 | -3 | 3 | 1 | -2 | 2 | 1 | -4 | -3 |
| 2 | -1 | -7 | 4 | -4 | 1 | -1 | 1 | -1 | 1 | -2 | 3 | 0.05 | -0.05 | 0.05 | -0.05 | 0.05 | -0.15 | 0.3 | -0.3 | -0.15 | 0.1 | 0.45 |
| 4 | 1 | -3 | -4 | 4 | -1 | 1 | -1 | 1 | -1 | 2 | -3 | -0.05 | 0.05 | -0.05 | 0.05 | -0.05 | 0.15 | -0.3 | 0.3 | 0.15 | -0.1 | -0.45 |
| 6 | -1 | 3 | -6 | -4 | 1 | -1 | 1 | -1 | 1 | -2 | 3 | 0.05 | -0.05 | 0.05 | -0.05 | 0.05 | -0.15 | 0.3 | -0.3 | -0.15 | 0.1 | 0.45 |
| 8 | 1 | -3 | 6 | -6 | -1 | 1 | -1 | 1 | -1 | 2 | -3 | -0.05 | 0.05 | -0.05 | 0.05 | -0.05 | 0.15 | -0.3 | 0.3 | 0.15 | -0.1 | -0.45 |
| 10 | -1 | 3 | -6 | 6 | -9 | -1 | 1 | -1 | 1 | -2 | 3 | 0.05 | -0.05 | 0.05 | -0.05 | 0.05 | -0.15 | 0.3 | -0.3 | -0.15 | 0.1 | 0.45 |
| 12 | 3 | -9 | 8 | -8 | 7 | -7 | -3 | 3 | -3 | 6 | 1 | -0.15 | 0.15 | -0.15 | 0.15 | -0.15 | 0.45 | -0.9 | 0.9 | 0.45 | -0.3 | -1.35 |
| 14 | -6 | 8 | -6 | 6 | -4 | 4 | -4 | -6 | 6 | -2 | -2 | 0.3 | -0.3 | 0.3 | -0.3 | 0.3 | -0.9 | 1.8 | -1.8 | -0.9 | 0.6 | 2.7 |
| 16 | 6 | -8 | 6 | -6 | 4 | -4 | 4 | -4 | -6 | 2 | 2 | -0.3 | 0.3 | -0.3 | 0.3 | -0.3 | 0.9 | -1.8 | 1.8 | 0.9 | -0.6 | -2.7 |
| 18 | 3 | 1 | -2 | 2 | -3 | 3 | -3 | 3 | -3 | -4 | 1 | -0.15 | 0.15 | -0.15 | 0.15 | -0.15 | 0.45 | -0.9 | 0.9 | 0.45 | -0.3 | -1.35 |
| 20 | -2 | 6 | -2 | 2 | 2 | -2 | 2 | -2 | 2 | -4 | -4 | 0.1 | -0.1 | 0.1 | -0.1 | 0.1 | -0.3 | 0.6 | -0.6 | -0.3 | 0.2 | 0.9 |
| 22 | -9 | 7 | -4 | 4 | -1 | 1 | -1 | 1 | -1 | 2 | -3 | 0.45 | -0.45 | 0.45 | -0.45 | 0.45 | -1.35 | 2.7 | -2.7 | -1.35 | 0.9 | 4.05 |

**The role of the anchor group.**

M-theory is valid when the core of a PAH is weakly connected to the electrodes. The DFT results below show the effect of increasing the strength of the coupling to the electrodes by replacing the linker group of figure 1 (main text) with a direct carbon-gold bond.

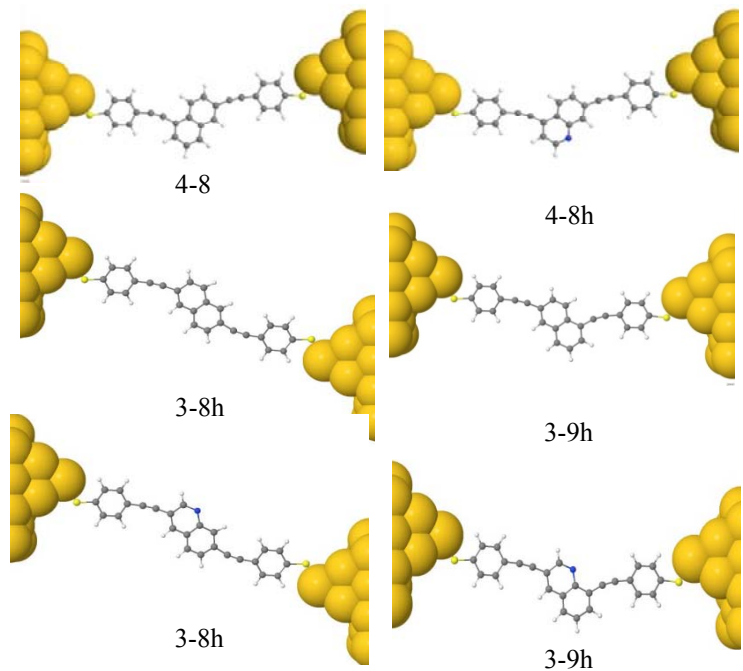

**Figure SI.1.** Relaxed structure for different connectivities of naphthalene with and without heteroatom connected to the electrode.

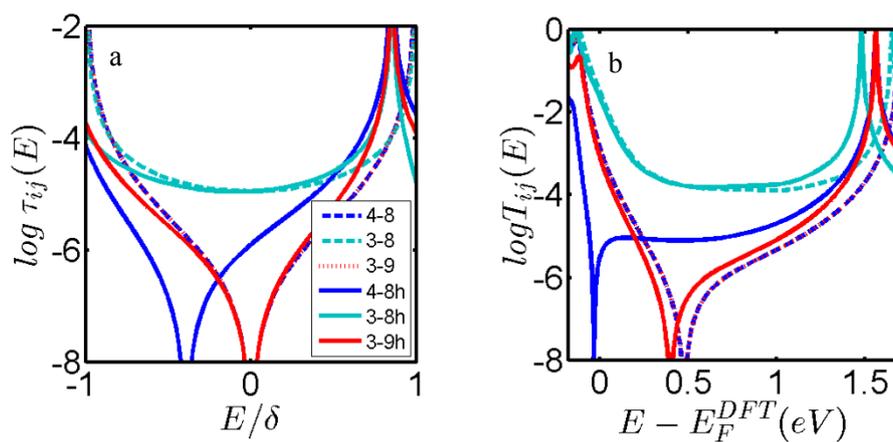

**Figure SI.2** (a) Core transmission coefficients $\tau_{i,j}(E)$ of parents (dashed lines) and daughters (solid lines) plotted against $E/\delta$, where $\delta$ is half of the HOMO-LUMO gap of the parental core; ie $\delta=0.62$ (b) *DFT* results for the corresponding the transmission coefficients $T_{ij}(E)$ of the naphthalene with the structures shown in figure SI.1.

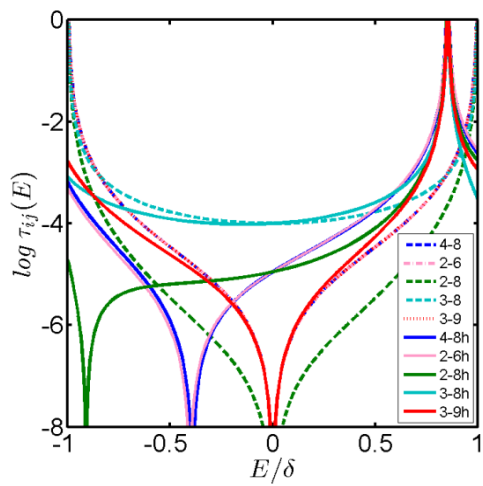

**Figure SI.3.** Core transmission coefficients $\tau_{i,j}(E)$ of naphthalene (dashed lines) and quinolone (solid lines).

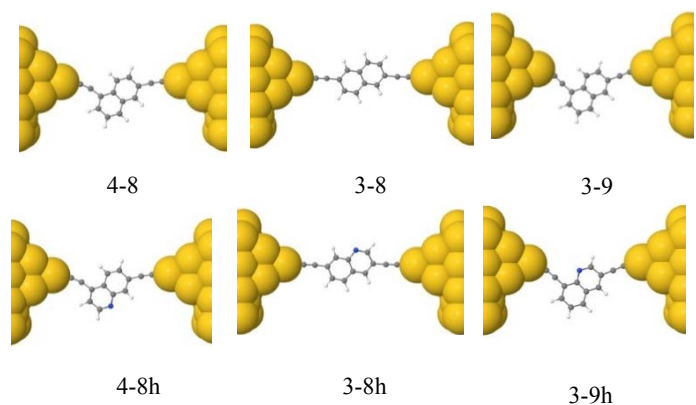

**Figure SI.4.** Relaxed structure for different connectivities of naphthalene with and without heteroatom connected to the electrode, with different anchor.

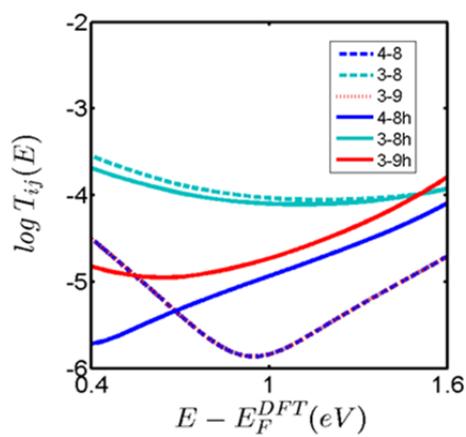

**Figure SI.5.** *DFT* results for the corresponding the transmission coefficients $T_{ij}(E)$ of the naphthalene with the structures shown in figure SI4.

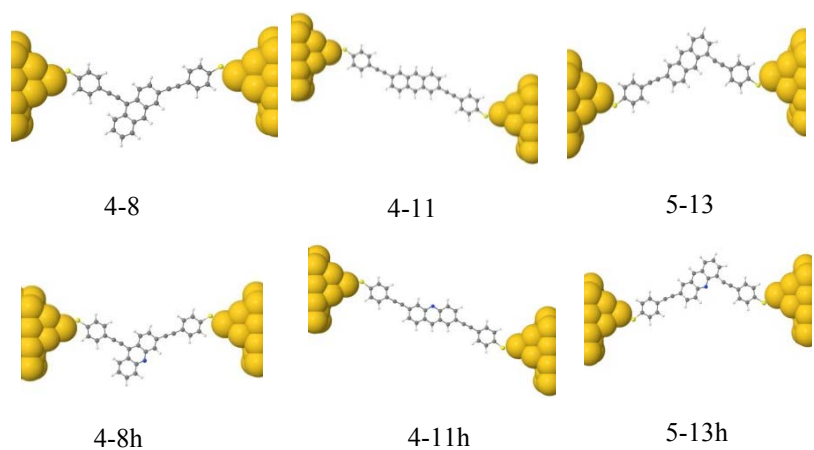

4-8    4-11    5-13

4-8h   4-11h   5-13h

**Figure SI.6.** Relaxed structure for different connectivities of anthracene with and without heteroatom connected to the electrode

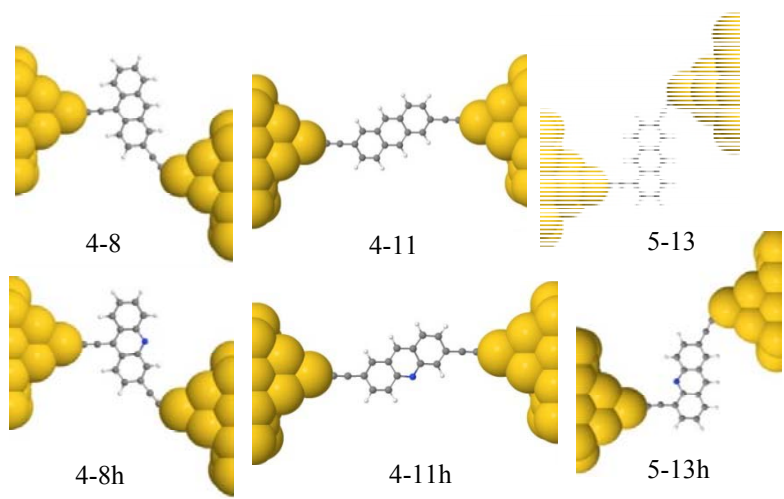

4-8    4-11    5-13

4-8h   4-11h   5-13h

**Figure SI.7.** Relaxed structure for different connectivities of anthracene with and without heteroatom connected to the electrode, with different anchor

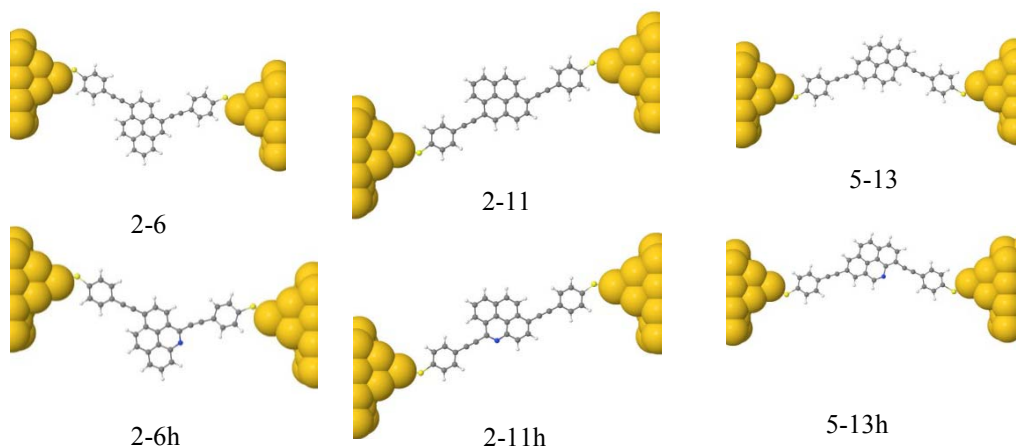

**Figure SI.8.** Relaxed structure for different connectivities of pyrene with and without heteroatom connected to the electrode

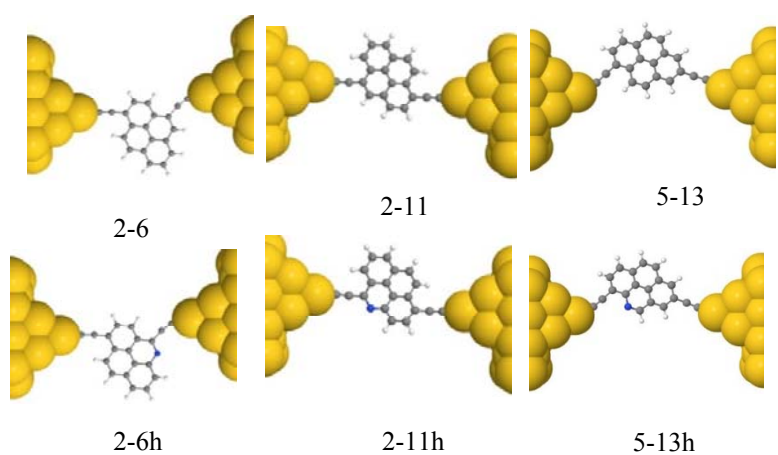

**Figure SI.9.** Relaxed structure for different connectivities of pyrene with and without heteroatom connected to the electrode, with different anchor

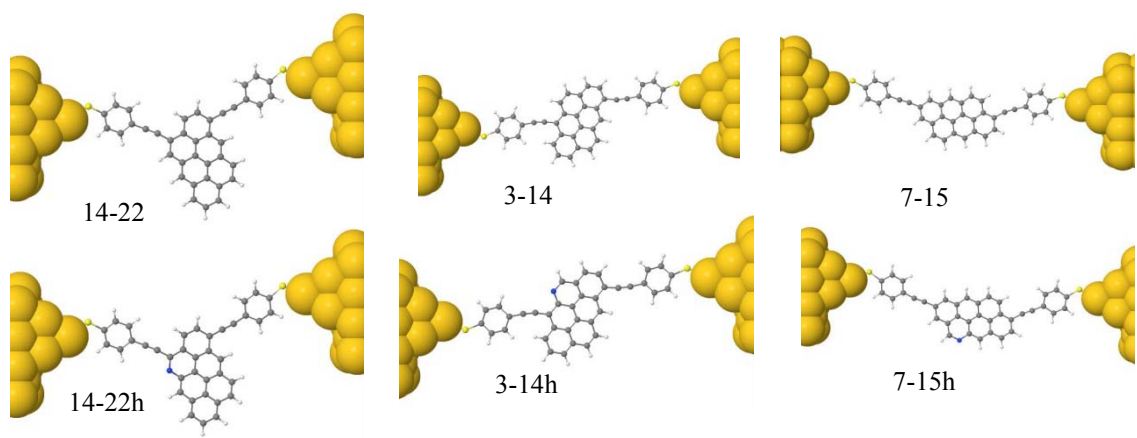

**Figure SI.10.** Relaxed structure for different connectivities of anthanthrene with and without heteroatom connected to the electrode

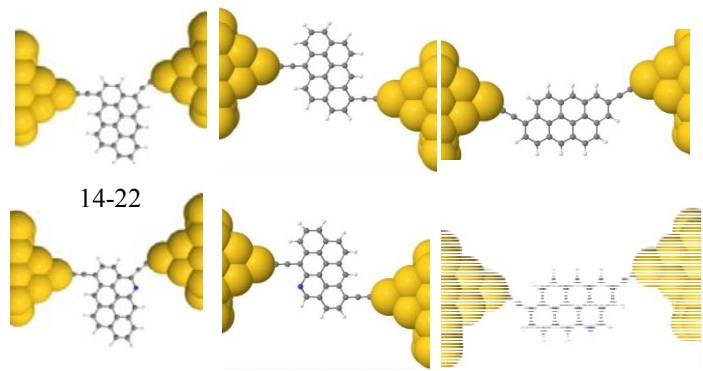

14-22

**Figure SI.11.** Relaxed structure for different connectivities of anthanthrene with and without heteroatom connected to the electrode, with different anchor

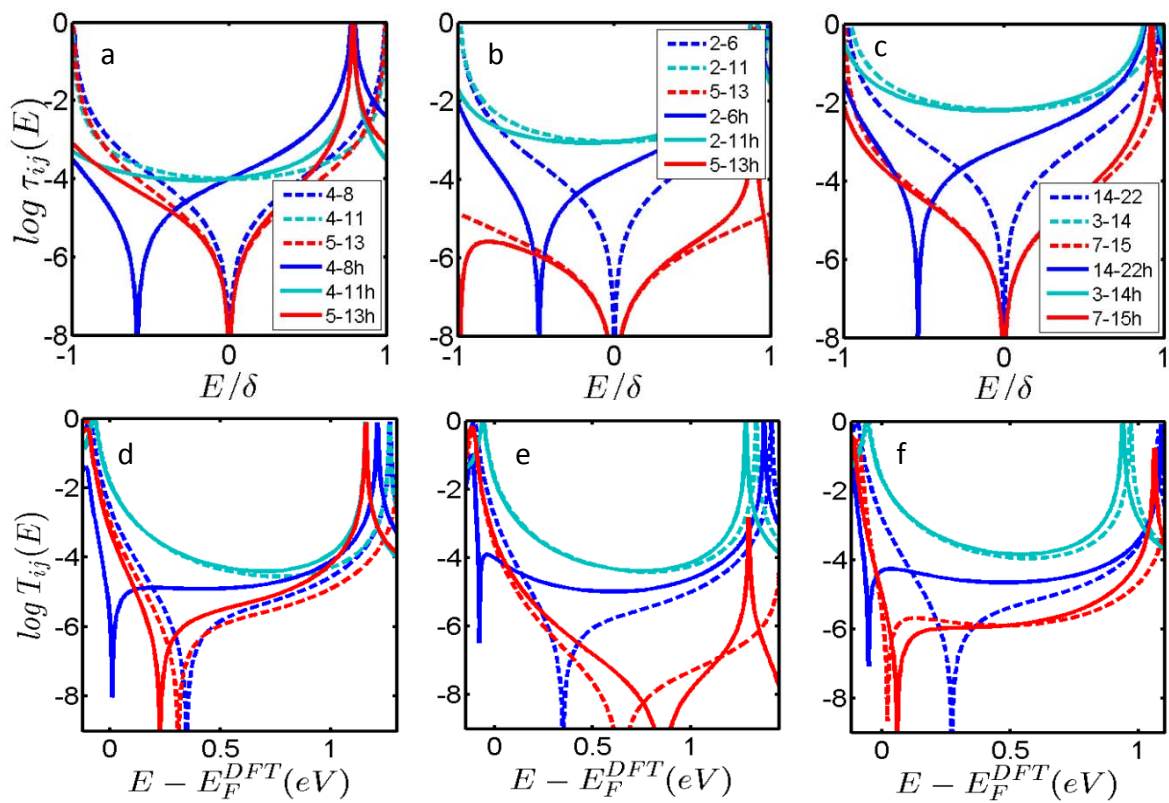

**Figure SI.12**. (a-c): Core transmissions of parents (dashed lines) and of daughters (solid lines), of (a) anthracene, (b) pyrene and (c) anthanthrene. (d-f): DFT-NEGF results for the zero-bias electrical conductances of daughters (solid lines) and parents (dashed Lines) of (d) anthracene, (e) pyrene and (f) anthanthrene., for structures shown in fig SI.6., SI.8, SI.10.

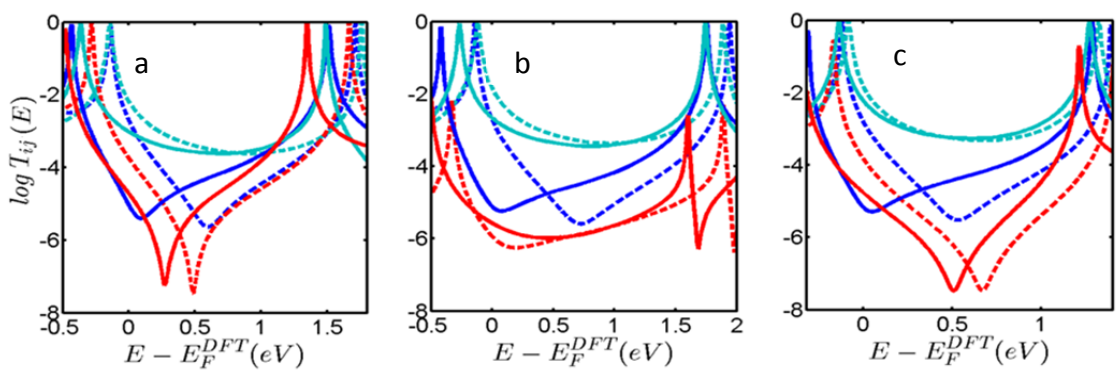

**Figure SI.13**. (a-c): DFT-NEGF results for the zero-bias electrical conductances of daughters (solid lines) and parents (dashed Lines) of (a) anthracene, (b) pyrene and (c) anthanthrene. .,for structures shown in fig SI.7., SI.9, SI.11.

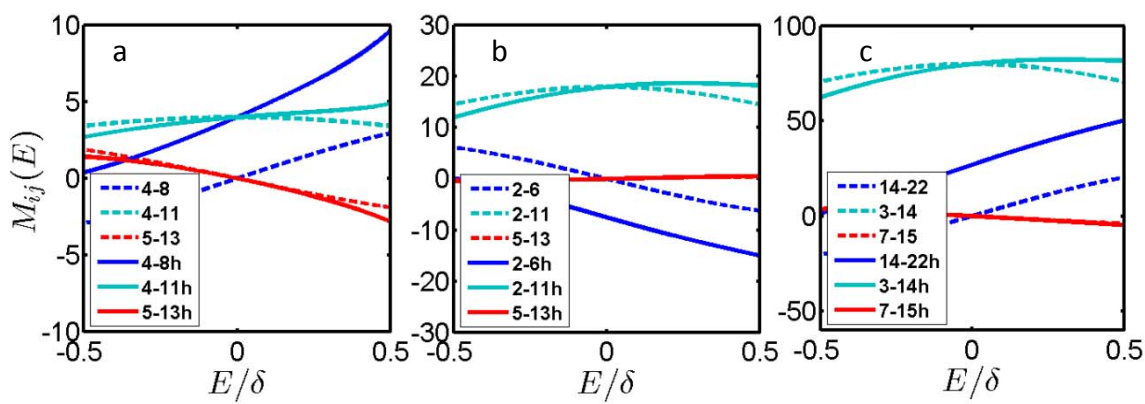

**Figure SI.14**. $M_{ij}(E)$ of the parent and daughter molecules of (a) anthracene, (b) pyrene and (c) anthanthrene